\documentclass[12pt]{article}

\usepackage{amsmath,amssymb,epsfig}
\usepackage{verbatim}

\def\beq{\begin{equation}} 
\def\eeq{\end{equation}} 
\newcommand{\ba}{\begin{array}}  
\newcommand{\ea}{\end{array}} 
\newcommand{\bea}{\begin{eqnarray}}  
\newcommand{\eea}{\end{eqnarray} }  
\newcommand{\bal}{\begin{align}}
\newcommand{\eal}{\end{align}}   
\def\bi{\begin{itemize}}  
\def\ei{\end{itemize}}  
\def\ben{\begin{enumerate}}  
\def\een{\end{enumerate}}  
\def\beq{\begin{equation}}  
\def\eeq{\end{equation}}  
\def\bc{\begin{center}}
\def\ec{\end{center}} 
 \def\bt{\begin{table}}
\def\et{\end{table}}  
 \def\btb{\begin{tabular}}
\def\etb{\end{tabular}}  
\newcommand{\bvec}{\left ( \ba{c}}
\newcommand{\evec}{\ea \right )}

\def\pa{\partial}

\def\simlt{\stackrel{<}{{}_\sim}}

\newcommand{\nn}{\nonumber \\}

\def\tev{\, {\rm TeV}}

\def\mkk{\, m_{\rm KK}}

\def\mass2{mass${}^2$}

\def\ra{\rangle}
\def\la{\langle}  

\def\pa{\partial}

\def\simlt{\stackrel{<}{{}_\sim}}

\newcommand{\ti}{\tilde}

  \def\eps{\epsilon}
  \newcommand{\lsim}{\;\raisebox{-0.9ex}{$\textstyle\stackrel{\textstyle<}
           {\sim}$}\;}

\begin{document}



\thispagestyle{empty}
\font\cmss=cmss10 \font\cmsss=cmss10 at 7pt

\begin{flushright}
ANL-HEP-PR-07-104 \\
CERN-PH-TH/2007-247\\
SPhT-T07/153\\
SU-4252-863 \\
UMD-PP-07-008\\
\end{flushright}

\vspace{5pt}

\begin{center}
{\Large \textbf
{KK Parity in Warped Extra Dimension}}
\end{center}

\vspace{10pt}
\begin{center}
{\large Kaustubh Agashe$\, ^{a, b}$, Adam Falkowski$\, ^{c, d}$,
Ian Low$\, ^{e, f, g}$, G\'eraldine Servant$\, ^{c, h}$}\vspace{15pt}

$^{a}$\textit{Department of Physics, 
Syracuse University, Syracuse, NY 13244, USA}
\\
$^{b}$\textit{Maryland Center for Fundamental Physics,
     Department of Physics,
     University of Maryland,
     College Park, MD 20742, USA}
\\
$^{c}$\textit{CERN, Theory Division, CH 1211, Geneva 23, Switzerland}
\\
$^{d}$\textit{Institute of Theoretical Physics, Warsaw University, Ho\.za 69, 00-681 Warsaw, Poland}
\\
$^{e}$\textit{Department of Physics and Astronomy,
University of California, Irvine, CA 92697, USA}
\\
$^{f}$\textit{Theory Group, HEP Division, Argonne National Laboratory, Argonne, IL 60439, USA}
\\
$^{g}$\textit{\mbox{Department of Physics and Astronomy, Northwestern University, Evanston, IL 60208, USA}}
\\
$^{h}$\textit{Service de Physique Th\'eorique, CEA Saclay, F91191, Gif-sur-Yvette, France}

\vskip .75cm
{\tt  kagashe@umd.edu, adam.falkowski@cern.ch,\\
ilow@northwestern.edu, geraldine.servant@cern.ch}

\end{center}

\vspace{2pt}
\begin{center}
\textbf{Abstract}
\end{center}
\vspace{5pt} { \noindent

We construct models with a Kaluza-Klein (KK) parity in a five-dimensional 
warped geometry, in an attempt to address the little hierarchy problem 
present in setups with bulk Standard Model fields.  
The lightest KK particle (LKP) is stable and can play the role of  dark matter. 
We consider the possibilities of gluing two identical slices of AdS$_5$ in 
either the UV (IR-UV-IR model) or the IR region (UV-IR-UV model) and 
discuss the  model-building issues as well as phenomenological properties in both cases.
In particular, we find that the UV-IR-UV model is not  gravitationally stable
and that additional mechanisms might be required in the
IR-UV-IR model to  address flavor issues.
Collider signals of the warped KK parity are different from either the 
conventional warped extra dimension without KK parity, in which the new particles are not necessarily 
pair-produced, or the KK parity in flat universal extra dimensions, where each KK level is
nearly degenerate in mass. Dark matter and collider properties of a TeV mass KK $Z$ gauge boson as the LKP are discussed. 
}

\vfill\eject
\noindent


\section{Introduction}

Solutions to the hierarchy problem of the Standard Model (SM) invoke new 
physics (NP) around the TeV scale to cut-off the quadratically divergent 
quantum corrections to the Higgs mass. Ideally, to avoid  too much 
fine-tuning, the lightest NP states should  be present already at the weak 
({\em sub}-TeV) scale.  However, NP induces higher-dimensional operators 
involving the SM particles which result in a tension with precision tests of 
the SM, in both the electroweak (EW) and the flavor sector. To be consistent 
with the EW precision tests, flavor-preserving operators generated by NP 
typically require the scale of NP to be larger than {\em a few} TeV 
\cite{Barbieri:1999tm} and are difficult to suppress by any known 
(approximate) symmetries of the SM\footnote{Exceptions include custodial 
isospin for the $T$ parameter \cite{Peskin:1991sw}.}.  This tension is called 
the little hierarchy problem. Besides, in the presence of $O(1)$ new sources 
of CP violation, the data on flavor violation in Kaon system requires the NP 
mass scale to be larger than several thousands TeV. However, it might be 
possible to address the latter constraints by suitable flavor symmetries.

A new symmetry at the TeV scale can ameliorate some of these constraints if at least the lightest
NP states, which a priori give the largest electroweak corrections, are charged 
under this symmetry while the SM particles are neutral \cite{Wudka:2003se,Cheng:2003ju}. 
In such a case, the charged NP states do not contribute at 
 tree level to
the operators constrained by the precision tests
since couplings of a single charged state
to SM particles are forbidden.
NP contributions from these states arise only at loop level. This  makes  sub-TeV NP states consistent with  EW precision data.
These NP states may then play the role of cutting-off the Higgs mass divergence  without any fine-tuning, thus avoiding the little hierarchy problem.
As a spin-off, the new symmetry implies the existence of a new stable particle that can be a dark matter candidate if it is electrically neutral and weakly interacting. 

The simplest possibility of a new symmetry at the TeV scale is a discrete $Z_2$ parity.
The classic example is $R$-parity in supersymmetry.
In little Higgs, the similar role is played by $T$-parity \cite{Cheng:2003ju,Cheng:2004yc} under which the new gauge bosons are charged. 
Yet another example is Kaluza-Klein (KK) parity \cite{Cheng:2002ab} in universal extra dimensions (UED) \cite{Appelquist:2000nn}. 
However, no explicit UV completions exist in the literature for the latter two scenarios, which by nature are effective theories below say 10 TeV. Moreover, all three of these frameworks do not 
address flavor violation issues which require detailed understanding of the possible 
UV completion or SUSY breaking mechanism.

The situation is quite different in the Randall--Sundrum (RS1) setup \cite{Randall:1999ee} based on 
a slice of AdS$_5$ in the sense that
both the Planck-weak and flavor hierarchies can
be addressed as follows.
%
%
Owing to the warped geometry, the 4D (or zero-mode) graviton is localized near the UV/Planck
brane which has a Planckian fundamental scale, 
whereas the Higgs sector can be localized near the IR/TeV brane where the cut-off is of order TeV. 
In this way the Planck-weak hierarchy is addressed. 
Based on the AdS/CFT correspondence \cite{Maldacena:1997re, Witten:1998qj}, RS1 is conjectured to be dual to 4D composite Higgs models \cite{Arkani-Hamed:2000ds, Rattazzi:2000hs, Contino:2003ve}.
%
%
In the original RS1 model, the entire SM (including the fermions
and gauge bosons) are assumed to be localized on the TeV brane.
However, it was subsequently realized
that, with the SM fermion \cite{gn, gp} and gauge fields
\cite{bulkgauge} propagating in the bulk,
such a framework not only solves the Planck-weak hierarchy, but can 
also address the flavor hierarchy.
The idea is that light SM fermions
(which are zero-modes of 5D fermions)  can be localized near the
UV brane, whereas the top quark is localized near the
IR brane, resulting in small and large couplings respectively
to the SM Higgs localized near the IR brane.
Moreover, the flavor problem 
(both from unknown physics at the cut-off and from the KK states)
is also under control
\cite{gp, Huber:2000ie} due to an analog of the GIM mechanism  or
approximate
flavor symmetries \cite{Agashe:2004cp},
even with a few TeV KK scale and despite
 the recent $B$-physics data \cite{NMFV}.

The versions of this framework studied so far do not
have a discrete symmetry analogous to KK parity in UED.
The constraints from EW precision tests
require the lightest gauge KK modes to be heavier than a few TeV,
provided suitable custodial symmetries are implemented
to suppress contributions to the $T$ parameter \cite{Agashe:2003zs}
and shift in coupling of $Z$ boson to $b_L$ \cite{Agashe:2006at}.
As mentioned above, a similar limit on
the KK mass scale arises
also from flavor violation: see references
\cite{others1, others2, Cacciapaglia:2007fw}
for other studies of these issues.
%
%
%
Thus, although the big (Planck-weak) 
hierarchy and flavor issues are addressed, the little hierarchy problem generically
persists in these models. 
Phenomenologically, the implication of the little hierarchy is that, if mass scales of the new physics 
are higher than $2 - 3$ TeV, the new particles would barely be reachable at the LHC especially if they
are not charged under the $SU(3)_c$ strong interaction \cite{kkgluon}. 

The goal of this paper is to implement an
analog of KK-parity of UED in a warped extra dimension, by requiring the warp factor
to be symmetric with respect to the mid-point of the extra dimension. In this construction, 
there are two towers in the KK decomposition
of a bulk field, namely, KK modes which are even and odd under the parity symmetry. 
The SM particles belong to the even towers.
The odd modes cannot have single couplings to the SM, therefore  they are allowed to be lighter than a TeV
without contradicting the precision EW constraints.
%
%
%
Although the primary focus of the present work is to ease out the experimental 
constraints and lower the mass scale
of the new particle, we will
argue that  these lightest odd modes
can cut off the quadratic divergences in the Higgs sector,
thus addressing the little hierarchy problem.
Furthermore, the lightest odd particle is stable and could be a WIMP, naturally
giving the correct dark matter abundance, like  in UED \cite{Servant:2002aq,Cheng:2002ej}.
The resulting
collider phenomenology is different from usual models
with a warped extra dimension. In particular, 
KK-odd particles have to be produced in pairs
and give missing
energy signals due to the decay chains ending in
lightest KK-odd particles.
%

The outline of the paper is as follows. In the next section
we present a brief review of KK number conservation and
KK parity in UED. In Section \ref{threesite} we discuss three-site moose toy models to understand
the relation between different warp factors and the low-energy KK spectrum of gauge bosons.
 In Section  \ref{IRUVIR} we consider gluing two
identical slices of AdS$_5$ in the UV region (the IR-UV-IR setup) and discuss the phenomenological 
features. Large brane-localized
terms are necessary in order to obtain the desired pattern for the spectra of gauge bosons. 
In this Section,
we present a model where the LKP is a KK $Z$ gauge boson and discuss the corresponding dark matter phenomenology. In Section
\ref{UVIRUV} we discuss briefly the alternative of gluing two slices of AdS$_5$ in the IR region (the UV-IR-UV
setup). Even though the
UV-IR-UV setup has certain nice phenomenological features, this setup is unstable gravitationally. In the last 
section we present our conclusions. Lastly in the appendices we give a
CFT interpretation of our setups, as well as some discussion on cutting off the Higgs quadratic-divergences using the lightest KK-odd gauge bosons.

\section{Mini-Review on UED}
\label{ued}

We begin by reviewing origins of the success of UED  \cite{Appelquist:2000nn} in fitting the precision electroweak measurements while 
allowing for KK masses well below 1 TeV, as
certain important features of UED have not been emphasized enough in the past, which nonetheless 
will become crucial when constructing models with KK parity in warped extra dimension.
Hence, this review of UED will serve as a guide in
model-building for the warped case.

In the framework of UED, the existence of KK parity requires very special conditions. 
In that setup, the extra dimension is an interval with a flat background geometry, and
KK parity is realized as a geometric reflection about the midpoint of the extra dimension.\footnote{
Strictly speaking, in Ref.~\cite{Cheng:2002ab}, KK-parity is defined as the reflection about the midpoint combined with the orbifold projection. However, one could instead work on the line interval without referring to the orbifold at all. We come back to this when discussing bulk fermion mass.} 
Alternatively, such an extra
dimension can be viewed as an orbifold $S^1/Z_2$, that is a compactified circle with a 
$Z_2$ orbifolding imposed. Before $Z_2$ orbifolding, the circle $S^1$ has a translational symmetry that
is manifested as a $U(1)$ symmetry in the 4D KK decomposition. Momentum in the fifth direction now becomes
quantized and each KK mode carries a conserved quantum number, the KK number, under the $U(1)$ symmetry. 
The translational symmetry along the circle is obviously broken by the $Z_2$ orbifolding, or, in other words, by the
orbifold fixed points, which can be thought of as boundaries or branes at the ends of the extra dimension. However,
it is clear that a discrete subgroup of the translation survives
(assuming that any interactions, whether
large or small, localized on the two branes
are equal), leading to the KK parity.

The picture of $S^1/Z_2$ orbifold makes it clear that KK parity has a larger parent symmetry, 
the KK number conservation, which is broken only by the interactions living on the branes at the ends of the 
interval. In the literature on UED models, it is usually assumed that the brane-localized 
interactions are symmetric with respect to the $Z_2$ reflection about the midpoint, so that KK parity is an exact symmetry. It is also assumed that they are
suppressed (loop-induced), implying that KK number is still an approximate symmetry.  These assumptions have very important phenomenological implications, as both KK parity and the approximate
KK number conservation are needed to evade precision electroweak constraints for UED models; KK parity eliminates  couplings of a single odd KK mode with the SM field, whereas the approximate 
KK number conservation
suppresses certain interactions among the even level KK modes, such as single coupling of the 2nd KK mode with 
the SM, which are not
forbidden by KK parity. In the end, both the odd and even KK modes are allowed to have masses well
below 1 TeV. If there were only KK parity and not the approximate KK number conservation, experimental
constraints would have required the 2nd KK mass to be higher than 2 - 3 TeV and, therefore, the compactification
scale to be around 1 TeV or higher (recall that in flat geometry KK modes are evenly spaced).

One should keep in mind that the flatness of profiles in UED is not natural and reflects the fact that electroweak 
symmetry breaking is not addressed but just postulated. A model of dynamical symmetry breaking in UED would typically 
spoil the flatness of the Higgs profile and constraints on the KK scale would have to be reexamined 
accordingly (a somewhat related discussion on the little hierarchy problem in UED is presented 
in \cite{Burdman:2006jj}). The virtue of UED is that mass scales of new particles are allowed to be very close to the 
electroweak scale at a few hundreds GeV, allowing for easy access at the LHC, even though the model 
addresses neither the Planck-weak nor the fermion mass hierarchy as it stands in 
the literature.

Since the KK number conservation which prevents the 2nd KK 
mode from giving large electroweak corrections has its origin in the flat background geometry in the extra dimension, 
it is clear  that it will be lost in a curved background. 
As a consequence, if we want to implement KK parity in a warped extra dimension, all the higher
even KK modes will have un-suppressed couplings with the SM and be required to be heavier than 2 - 3 TeV,
as dictated by the model-independent analysis.
On the other hand, all KK modes odd under KK parity still need to couple in pairs 
to the SM and can only 
contribute to electroweak observables at the loop level. 

Contrary to UED, a warped extra dimension allows us to investigate
various UV sensitive questions such as the Planck-weak hierarchy problem.
However, before going into a full-fledged extra-dimensional setup, it is instructive to consider a 
low-energy effective description involving only up to the 2nd KK mode of the gauge boson. Since higher KK modes 
might be too heavy to be accessible at the LHC, such an effective theory may be all that matters at the collider
experiments and we present this discussion in the next section.
\section{Three-site Toy Model}
\label{threesite}

\begin{figure}[t]
\begin{center}
\includegraphics[width=7cm]{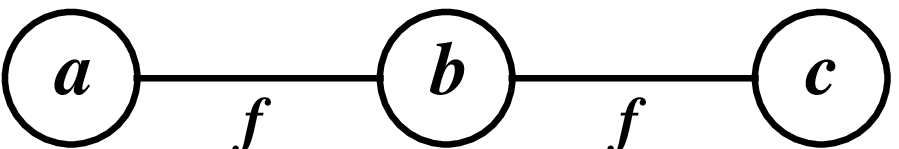}
\caption{\it Three-site deconstruction of the warped extra-dimension. }
\label{threesitefig}
\end{center}
\end{figure}

  In essence, the low-energy effective
theory amounts to a three-site deconstruction \cite{Arkani-Hamed:2001ca,Hill:2000mu} of the 
warped extra dimension; see Fig.~\ref{threesitefig}. The gauge symmetry at each site is denoted $G_i$, $i=a, b, c$, with corresponding gauge
bosons $A^{(i)}_\mu$. In general, the gauge coupling constants and the decay constants can be
 different at each site, unlike in the case of flat
background geometry. However, the KK parity, which in the current setup is the geometric reflection 
$a \leftrightarrow c$, ensures that the gauge couplings on the two boundary sites as well as the two
decay constants are equal. 
It is then straightforward to
work out the low-energy spectrum of the three-site model. Defining the zero-mode gauge coupling to be
\begin{equation}
\frac1{g_0^2} = \frac2{g_a^2} +\frac1{g_b^2},
\end{equation}
the mass eigenvalues and eigenstates are 
\begin{equation}
m_0 = 0, \quad  m_{1_-} = g_a f ,  \quad m_{1_+}= \sqrt{g_a^2 + 4g_b^2} f, 
\end{equation}
and
\begin{eqnarray}
\label{threesiteeig}
A_\mu^{(0)} &=& \frac{g_0}{g_a} \left(A^{(a)}_\mu + A^{(c)}_\mu\right) + \frac{g_0}{g_b} A^{(b)}_\mu, \nonumber \\
A_\mu^{(-)} &=& \frac1{\sqrt{2}} \left(A^{(a)}_\mu - A^{(c)}_\mu\right),  \\
A_\mu^{(+)} &=& \frac{g_0}{\sqrt{2} g_b} \left(A^{(a)}_\mu + A^{(c)}_\mu\right) -
         \frac{\sqrt{2}g_0}{g_a} A^{(b)}_\mu. \nonumber
\end{eqnarray}
As a first check, we see both the zero-th and second KK modes are even under the KK parity, 
$a\leftrightarrow c$, whereas the first KK mode is odd. Furthermore, we see that the KK masses are controlled by
the gauge couplings on the boundary and the middle sites. Two particular limits we are interested in are 
\begin{eqnarray}
\frac{g_a}{g_b} \gg 1  &\Rightarrow& \frac{m_{1_-}}{m_{1_+}} \approx 1 - \frac{2g_b^2}{g_a^2} 
\approx 1; \\
\frac{g_a}{g_b} \ll 1  &\Rightarrow& \frac{m_{1_-}}{m_{1_+}} \approx \frac12 \frac{g_a}{g_b} 
\ll 1.
\end{eqnarray}
In the first case when the gauge coupling at the boundary is much larger than the coupling in the middle, the
two massive KK modes are roughly degenerate. In the other case when the coupling in the middle site is 
much larger than the coupling on the boundary, the odd KK mode is ``anomalously'' light compared to the 2nd KK
mode and there can be a sizeable hierarchy between the two KK modes. 

If we view the three-site model as deconstruction of the warped extra dimension, the two limiting cases actually
correspond to two opposite types of warped geometries. It is useful to observe that the massless wave function
in Eq.~(\ref{threesiteeig}) is always localized  where the gauge coupling is smaller; the wave
function is localized near
the boundary sites if $g_a\ll g_b$ and the middle if $g_a\gg g_b$. The massive modes, on the other hand, are
localized away from where the gauge coupling is small due to orthogonality conditions. In models with warped 
extra dimension, it is well-known that 
%
%
the massive modes are localized toward the IR region \cite{bulkgauge, gp}. 
%
%
The above observation suggests that, in the case of $g_a/g_b \gg 1$, 
the two boundary sites mimic the IR region whereas the middle site is the UV region. In other words, it would
correspond to an IR-UV-IR warp factor which is symmetric with respect to the reflection about the middle site. 
This geometric $Z_2$ symmetry again serves as the source of the KK parity in our setup of warped extra dimension.
The other case of $g_a/g_b \ll 1$ then corresponds to the opposite
situation in which the two boundaries correspond to the UV region.
 This is the UV-IR-UV setup.

Another way of understanding the same statement is through the fact that the
smaller gauge coupling at a particular site implies that the strong coupling scale (the Landau pole) of the 
gauge theory is higher. In the warped extra dimension, a local cutoff, where the theory becomes strongly coupled, at a particular location, is determined by the warp factor at that
point, so that the UV region has a higher local cutoff than the IR region. 
Then, we arrive at the same conclusion that,  without additional contribution
from brane-localized kinetic terms, that could generate some hierarchy between the boundary and bulk couplings, the IR-UV-IR setup will lead to (almost) degenerate first (odd) and second
(even) KK gauge bosons, whereas in the UV-IR-UV setup there could be a (little) hierarchy between the odd and
even KK modes.


\section{IR-UV-IR Model}
\label{IRUVIR}

In order to obtain a warp factor which is symmetric with respect to reflection 
about the midpoint of the extra dimension, we consider joining two slices of 
AdS$_5$ since a single slice does not have such a symmetry.\footnote{Setups 
with more than one slice of AdS$_5$ space have been discussed in 
Refs.~\cite{Cacciapaglia:2005pa,Cacciapaglia:2006tg}, even though a symmetric 
warp factor was not considered.} Clearly there are two distinct ways to do 
this. We can glue the two slices either in the UV or in the IR region. We 
begin with the first possibility, labeled as the IR-UV-IR model.

The metric of the 5D  background spacetime resulting from  gluing two AdS$_5$ slices at the UV brane is
\begin{eqnarray}
ds^2 = dy^2 + a^2(|y|)d x^2,
\end{eqnarray}
where $y \in  [-L,L]$ is the extra dimension and
$a(y) =  e^{ - k y }$  is the warp factor. In order to obtain a Planck-Weak hierarchy, we choose $kL\sim 30$. Notice that in the conventional models of single slice AdS$_5$ the extra dimension is only $y \in [0,L]$.
This constitutes a solution of the 5D Einstein's equations in the presence of a negative bulk cosmological constant, a positive tension midpoint (the UV brane) and two IR branes with equal negative tensions. 
In such a setup the kinetic term of the massless radion has the correct sign and the radius
can be stabilized (i.e., radion made massive) by a suitable mechanism as usual \cite{Goldberger:1999uk}.
Therefore,  there  are no problematic stability issues associated  with the IR-UV-IR model 
as opposed to the  UV-IR-UV model described in the next section.

We assume the  $Z_2$ parity in interchanging $y \to - y$ is an exact symmetry
of the 5D theory.\footnote{%
If the Chern-Simons term is present in 5D, it is necessarily odd under $Z_2$. 
This would affect stability of the dark matter particle, as pointed out in 
\cite{Hill:2007zv}.  The Chern-Simons term could  arise in the presence of 
brane-localized anomalies \cite{ArkaniHamed:2001is}. In the following we will 
assume that all brane-localized anomalies cancel and no Chern-Simons terms are 
present.    } In such a case, the eigenmodes can be divided into two classes 
with different symmetry properties: 
 even modes, whose profiles are symmetric under reflection around the mid-point,
and  odd modes with anti-symmetric profiles. 
Obviously, the even and odd profiles are orthogonal to each other on the $[-L,L]$ interval. 
As long as the action respects the exact $Z_2$ symmetry, the odd modes can only couple in pairs to the even modes
in the KK decomposition and the low-energy, four-dimensional effective theory has the KK parity we desire for.

By continuity, the odd modes satisfy the Dirichlet boundary conditions, henceforth denoted by $(-)$, at the UV brane. 
Similarly, the even modes have Neumann $(+)$ boundary conditions (in the presence of UV brane localized terms mixed 
boundary conditions  (BC) arise). 
This observation suggests a useful description of the model by referring to the spectrum of a {\em single} slice of AdS$_5$. 
Namely, the spectrum of a bulk field in the $Z_2$-symmetric model contains two single-slice KK towers 
corresponding to $(+)$ and  $(-)$ boundary conditions in the UV. 
For example, a bulk field with Neumann boundary conditions in IR would have both $(++)$ and $(-+)$ towers, 
where the first (second) sign is the BC on the UV (IR) brane. 
Note however that the physical volume of the extra dimension in our setup is twice as large as in the single-slice
description, which affects the normalization of the wave functions.

At this point we already have a model combining warped geometry and KK parity. 
This is not the end of the story, however. 
As outlined in the previous section, one of our objectives is to obtain fairly light odd $(-+)$ modes 
(we would like
these modes to cut off the quadratic divergences in the Higgs mass) and sufficiently heavy $(++)$ 
KK modes (so as to evade tight constraints from  the precision electroweak tests). 
Unfortunately, in the simplest version with no brane kinetic terms the even and odd KK modes are quite degenerate,
as exemplified in the three-site model in the previous section. 
Both modes have masses of  order $m_{\rm KK} = k e^{ - k L }$ with their relative splitting being of 
order $\sim 1 /( k L ) \sim 1/30$. Another way to understand the degeneracy 
is that the AdS geometry 
localizes KK modes near the IR brane so that their spectrum is little sensitive to the UV brane boundary conditions.
As we discuss next, a splitting between even and odd {\em gauge} KK modes can be obtained with  very large IR brane kinetic terms 
(BKT), which in turn have important implications on the strong coupling scale in the 5D setup.

\subsection{Gauge bosons with large IR brane kinetic terms}

%
%
We consider the spectrum of {\em gauge} KK modes. A similar
analysis can be performed for other fields.
We follow the notation of Ref.~\cite{Carena:2002dz}
for the BKT's (see also Ref.~\cite{Davoudiasl:2002ua}).
The 5D action is
\begin{eqnarray}
\label{5daction}
S & = & - \int d^4x \int_{-L}^{L} dy \sqrt{- g}\
\frac1{4g_5^2} \Big[ F^{ M N } F_{ M N }+ 2 r_{ UV } F^{ \mu \nu } F_{ \mu \nu }  \delta ( y ) \nonumber \\
&& \hspace{5cm}
+  2 r_{ IR } F^{ \mu \nu } F_{ \mu \nu }  \delta ( y - L )
+ 2 r_{ IR } F^{ \mu \nu } F_{ \mu \nu }  \delta ( y + L )
\Big],
\end{eqnarray}
where $g$ is the determinant of the metric, and capital Latin letters 
$M,N=0,1,2,3,5$ refer to the 5D coordinates, whereas lower case Greek letters 
$\mu,\nu=0,1,2,3$ refer only to the four uncompactified directions. The 
strengths of the BKT on the two boundary IR branes are required to be equal by 
the $Z_2$ symmetry. Furthermore, each delta function on the boundary brane 
contributes only a factor of 1/2 when performing the $y$ integration.
%
%
%
%
%
Choosing the gauge $A_5=0$, we perform the KK decomposition by expanding 
\begin{equation}
A_\mu(x,y) = \sum_{n} A_{\mu, n}(x) f_{n}(y),
\end{equation}
where the bulk wave function $f_n(y)$ satisfies
\begin{eqnarray}
\label{5deom}
&&\partial_y\left[e^{-2k|y|}\partial_y f_n(y)\right] \nonumber \\
&& \hspace{2cm}   + m_n^2 \left[1+2r_{UV}\delta(y)
         +2r_{IR}\delta(y-L) + 2r_{IR}\delta(y+L)\right]f_n(y) = 0, \\
&& \label{5dnorm} 
  \frac1{g_5^2} \int_{-L}^{L} dy
   \left[1+2r_{UV}\delta(y)
         +2r_{IR}\delta(y-L) + 2r_{IR}\delta(y+L)\right]f_n^2(y) = 1.
\end{eqnarray} 
The $Z_2$ symmetry, $y\leftrightarrow -y$, inherited from the 5D action implies that 
bulk profiles are either even or odd under the reflection in the $y-$direction,
$f_n(y) = \pm f_n(-y)$. Therefore, we could rewrite the KK decomposition as
\beq
A_\mu(x,y) =\sum_{n_+, n_-} A_{\mu,n_+}(x) f_{n_+}(|y|) + A_{\mu,n_-}(x) \eps(y) f_{n_-}(|y|)
\eeq 
where  $f_{n_+}$ and $f_{n_{-}}$ are the even and odd modes, respectively, and $\epsilon(y)$ is
+1 $(-1)$ for $y>0$ ($y<0$).

Because of the warp factor $\exp(-2k|y|)$ in the equation of motion, one solves Eq.~(\ref{5deom})
separately for $y>0$ and $y<0$, imposes Neumann boundary conditions (mixed boundary conditions, in the presence of IR BKTs) at $y=\pm L$ to ensure a 
massless zero mode, and matches the solutions at $y=0$ as implied by the delta functions
in Eq.~(\ref{5deom}). When $r_{UV}=0$, the ``continuity conditions'' at $y=0$ are simply
\begin{equation}
f_{n_-}(0) = 0, \quad \partial_y f_{n_+}(0) = 0.
\end{equation}
As emphasized earlier, 
the above equation shows that a single bulk field in the IR-UV-IR setup would encompass modes that 
have both Dirichlet and Neumann boundary conditions on the UV brane, and we can simply ``borrow''
the results from the single-slice AdS$_5$ model by considering both types of boundary
conditions. In fact, using the $Z_2$ reflection symmetry, the 5D action in the IR-UV-IR 
setup in Eq.~(\ref{5daction}) can be re-written as
\begin{equation}
\label{new5daction}
S= -  \int dx^4 \int_{0}^{L} dy \sqrt{- g}\
\frac1{4\tilde{g}_5^2} \Big[ F^{ M N } F_{ M N }+ 2 r_{ UV } F^{ \mu \nu } F_{ \mu \nu }  \delta ( y ) 
+  2 r_{ IR } F^{ \mu \nu } F_{ \mu \nu }  \delta ( y - L ) \Big],
\end{equation}
where the integration in $y$ is only from 0 to $L$ and $\tilde{g}_5^2=g_5^2/2$. It is then clear 
that this is the 5D action 
of a single-slice AdS$_5$ with a re-defined 5D gauge coupling $\tilde{g}_5 = g_5/\sqrt{2}$, where
the factor of $\sqrt{2}$ represents the fact that the physical volume in the $y$-direction is 
actually twice as large as being integrated in Eq.~(\ref{new5daction}). Now it is straightforward
to construct the solutions to Eqs.~(\ref{5deom}) and (\ref{5dnorm}) by considering the equations 
of motion in the single-slice setup with $y\in [0,L]$:
\begin{eqnarray}
&& \pa_y (e^{-2 k y} \pa_y f_n) + m_n^2 f_n = 0 \\
&& \label{new5dnorm}
   \frac1{\tilde{g}_5^2} \int_0^L dy \left[1+2r_{UV}\delta(y)
         +2r_{IR}\delta(y-L)\right]f_n^2(y) = 1.
\end{eqnarray}     
and the boundary conditions
\bea
 e^{-2 k L} \pa_y f_{n_\pm}(L)  &=& m_{n_\pm}^2 r_{IR} f_{n_\pm}(L)
\\ 
\pa_y f_{n_+}(0)  &=& - m_{n_+}^2 r_{UV} f_{n_+}(0) \\
f_{n_-}(0) &=& 0
\eea 
The normalization in Eq.~(\ref{new5dnorm}) is consistent with Eq.~(\ref{5dnorm}) after taking 
into account $\tilde{g}_5 = g_5/\sqrt{2}$. 

The spectrum of the gauge boson in the IR-UV-IR setup now consists of two interlacing towers of 
modes, using the language of the single-slice model: the $(++)$ tower, which is KK-even, 
and the $(-+)$ tower, which is KK-odd. 
A massless mode in the $(++)$ tower always exists, irrespectively how large the BKTs are. 
We also have two towers of (roughly) equal-spaced KK modes starting at $\sim m_{\rm KK} = k e^{-k L}$. 
In addition, each tower has a parametrically lighter massive state.   
For $r_{IR} \gg 1/k$ we find the approximate expression  
\begin{eqnarray}
m_{1_-}^2 &\approx& \frac{2}{k r_{IR}} m_{\rm KK}^2  
\label{gaugeodd}
\\ 
m_{1_+}^2 &\approx& 
\frac{r_{UV} + r_{IR} + L}{r_{UV} + L} \frac{2}{k r_{IR}}  m_{\rm KK}^2  
\end{eqnarray}
As we can see, the lightest KK mode in each tower has its mass suppressed with respect to $m_{\rm KK}$. 
Of more importance to us is that we  can split the lightest even and odd modes. 
The ratio is  
\begin{equation}
\label{eoratio}
\frac{m_{1_+}}{m_{1_-}} \approx 
\sqrt{1 + \frac{r_{IR}}{r_{UV} + L}}
\end{equation}
Let us consider the effects of UV and IR BKT's in turn.
As mentioned earlier, in the absence of BKT's,
the even and odd modes are quite degenerate since
they are localized away from the UV brane so
they are insensitive to the different
BC's there (and BC's on IR brane are the same).
It is clear that very large UV BKT's, which affect only $(++)$ modes, could compensate the small UV brane wavefunction
and modify the spectrum of $(++)$ modes relative to $(-+)$ ones. 
However, we see from  Eq.~(\ref{eoratio}) that
for fixed IR BKT's, UV BKT's in fact tend to {\em reduce} the splitting
between even and odd KK modes.
It turns out that positive BKTs tend to repel massive KK modes away from the brane 
\cite{Carena:2002dz,Davoudiasl:2002ua} so that very large UV BKTs will 
effectively convert $(+)$ BC on the UV brane into $(-)$ BC, i.e., make the 2 towers even more degenerate.
Negative $r_{UV}$ increases the mass splitting, but it also leads to the appearance of a ghost 
(or  the Landau pole in the UV brane propagator) at the intermediate scale  $\sim k e^{- k |r_{UV}|}$.  
We cannot obtain a sizable splitting this way without lowering the UV brane cut-off very much. 
In the following we will set UV BKTs to be small or zero since they do not give the desired effects.

Consider next the effect of positive IR BKT's. In the absence of BKT's,
even and odd KK modes are localized near the IR brane. 
The BC (hence
the wavefunction) being the same on the IR brane for the even and odd towers, we might expect
the effect of IR BKT's on the two towers 
to be similar and therefore not lead to mass splitting. 
However, large positive IR BKTs tend to repel the massive wave functions away from
the IR brane, pushing them toward the  UV brane. In this
case, the spectrum would then become more sensitive to the BCs on the UV brane
[which are different for the $(++)$ and $(-+)$ modes], hence lead to a larger splitting
between the two modes.
%
%
However, to actually end up repelling the KK modes
away from
the IR brane, the BKT's have to overcome the ``pressure'' from
AdS geometry to localize KK's near the IR brane.
Only very large IR BKT's, $kr_{ IR }   \gg k L$, lead to
a large splitting between even and odd modes. To be precise:
\begin{eqnarray}
\frac{ m_{1_+} }{ m_{1_-} } & \sim & \sqrt{\frac{ k r_{ IR } }{ k L }}
\label{ratio}
\end{eqnarray}
The need for such a size of IR BKT's can be understood using the idea of the 
holographic RG flow. As explained in Ref. \cite{holoRG}, moving the UV brane 
by the infinitesimal proper distance $\eps$ toward the IR brane induces a 
brane kinetic term on the UV brane with a coefficient  $\propto \eps / g_5^2$. 
Moving the UV brane very close the IR brane we find that AdS without any brane 
kinetic terms is equivalent to flat space  with large brane kinetic terms  
$\sim L / g_5^2$ on one brane. Now, in the AdS model with large {\em IR} BKT's 
(but no UV BKT's to begin with), there is a competition between the UV brane 
terms {\em induced} via the holographic RG (which repel KK modes away from the 
UV brane) and the IR brane terms (which repel KK modes away from the IR brane) 
-- clearly the latter ``win'' for $r_{IR} \gg L$. Because of that repulsion 
away from the IR brane, the {\em even} gauge KK spectrum  is effectively given 
by $(+-)$ (in addition to a zero-mode which is effectively localized near the 
IR brane).  With such boundary conditions, there is the tower of KK modes 
starting at $\mkk$. In addition, there is a light mode whose mass is 
parametrically suppressed with respect to the  KK scale, $m_{1+} \sim  \mkk/(k 
L)^{1/2}$, as is well known from analysis of Higgsless models in AdS$_5$ 
\cite{higgsless}. Its mass is set by the zero-mode coupling in absence of 
large BKTs which is the origin of the $(k L)^{1/2}$.  This feature follows 
from the fact that this is a would-be zero mode (it would have been a 
zero-mode were it not for the effectively Dirichlet boundary condition on the 
IR brane) and its profile is almost flat except near the IR brane where  it is 
suppressed (see below). Similarly, the odd gauge KK spectrum is   effectively  
$(--)$, plus a would-be zero-mode localized near the IR brane.
Moreover, the fifth component $A_5$ has effectively $(++)$ BC, which yields a massless scalar mode that marries the would-be vector zero-mode. 
As a  consequence, the vector mass is set by the IR brane-localized gauge coupling and thus the suppression factor
in this mass is $(k r_{IR})^{1/2}$.

\begin{figure}[t]
\begin{center}
\includegraphics[width=5.3cm]{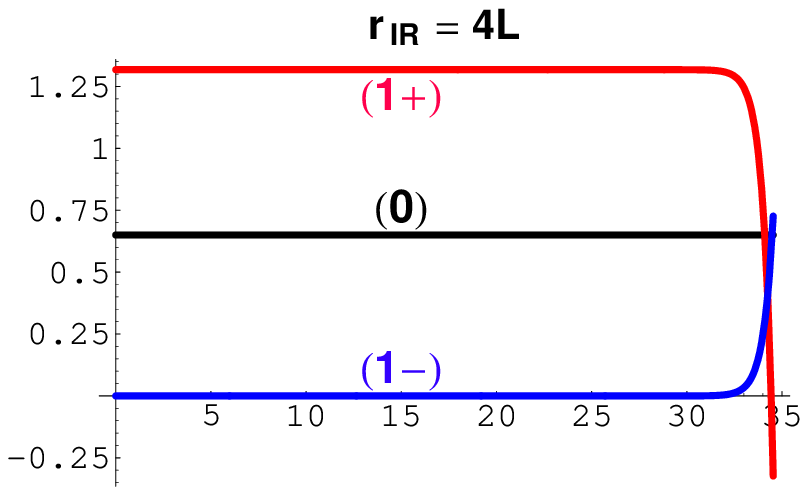}
\includegraphics[width=5.3cm]{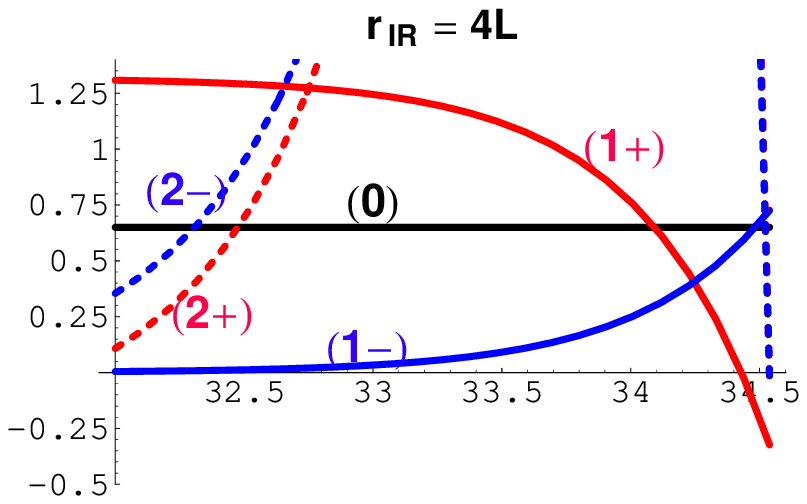}
\includegraphics[width=5.3cm]{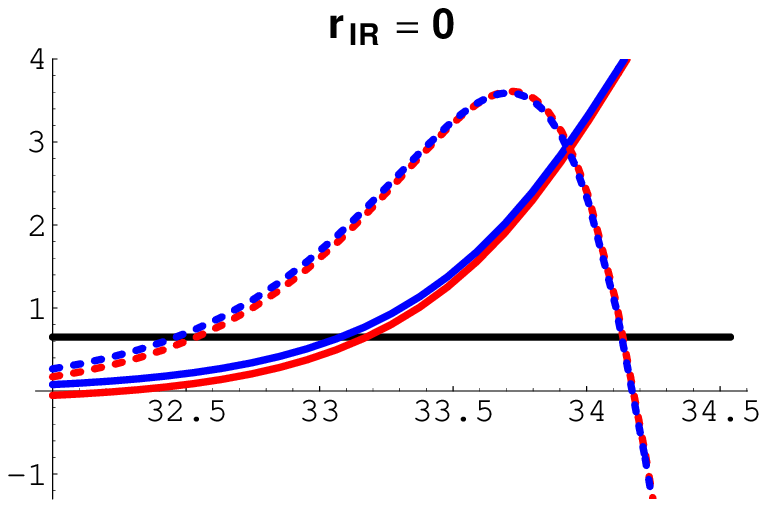}
\caption{\it Left: Gauge boson wave functions along extra dimension for first even $(1_+)$ mode (red), 
first odd $(1_-)$ mode (blue) and zero $(0)$ mode (black). 
Middle: Profiles are zoomed near the IR brane and we added in dashed lines the 
level-2 KK modes for comparison. Right: Same as middle plot but switching off the IR BKT. 
The first odd mode is more strongly coupled to the IR brane while the first even mode is less 
suppressed than in the case with IR BKT. }
\label{profiles}
\end{center}
\end{figure}

The profile of the lightest modes can be approximated by 
\begin{eqnarray}
f_0(y) &\approx & \frac{\tilde{g}_5}{\sqrt{r_{IR}}} 
\\
f_{1-}(y) &\approx& \frac{\tilde{g}_5}{e^{2 k L}\sqrt{r_{IR}}}  \left (e^{2 k y} - 1 \right )  
\\ 
f_{1+}(y) &\approx& \frac{\tilde{g}_5}{\sqrt{L + r_{UV}}} \left ( 1 -  
          \frac{1}{2k} m_{1_+}^2 (y + r_{UV}) e^{2 k y} \right ) .
\end{eqnarray}
It is important to remember that the wave functions here are written in terms of the ``re-defined''
5D gauge coupling $\tilde{g}_5$ in the single-slice AdS$_5$ action in Eq.~(\ref{new5daction}).
In the original formulation of IR-UV-IR setup in Eq.~(\ref{5daction}), $g_5= \sqrt{2} \tilde{g}_5$,
which would result in a suppression factor of $1/\sqrt{2}$  in the wave functions and account for the fact that
the physical volume in the extra dimension in the IR-UV-IR setup is twice as large as in the single-slice
AdS$_5$. Here we see that the zero-mode is flat and 
its normalization dominated by the IR BKT
so that it is {\em effectively} localized near IR brane. The zero mode gauge coupling, one of the low-energy observables,
is related to the 5D gauge coupling by $g_0 \approx  g_5/\sqrt{2r_{IR}}$. Again this is different from the
case of a single-slice AdS$_5$ setup by the volume factor.
As illustrated in Fig.~\ref{profiles}, 
the first odd mode is peaked at the IR brane, while the first even mode is almost flat everywhere except when it
is near the IR brane where the wave function is suppressed. From the profiles of the wave functions one sees that,
for $kL \gg 1$, the first odd mode couples to the IR brane with a similar strength to the zero mode,  
$f_0(L) \approx f_{1-}(L) \approx g_5/{\sqrt{2r_{IR}}}$, whereas its coupling to the UV brane is zero.
On the other hand, comparing to zero mode, the even mode coupling to the IR brane is suppressed,
\beq
\label{e.ireven}
\frac{f_{1+}(L)}{f_0(L)} 
\approx \sqrt{\frac{r_{UV} + L}{r_{IR}}} 
\approx 
\frac{m_{1_-}}{m_{1_+}},
\eeq   
while its coupling to the UV brane is {\em enhanced} (we will use
this fact later on): 
\beq
\label{crat}
\frac{f_{1+}(0)}{f_0(0)} \approx \sqrt{ \frac{r_{IR}}{r_{UV} + L} }
\approx \frac{m_{1_+}}{m_{1_-}}.
\eeq 
Such behaviors for the first KK-even mode have been observed in Refs.~\cite{Carena:2002dz,Davoudiasl:2002ua}
and can be understood from the repulsion of the 
corresponding wave functions
away from the branes due to the BKT's.
On the other hand, 
the zero-modes (including
the lightest odd mode which corresponds to
a ``would-be'' zero-mode) are not similarly repelled.
In the IR-UV-IR setup we found that the enhancement and suppression 
of the coupling of lightest even KK mode
to the UV and IR branes, respectively, is correlated with the mass splitting 
$m_{1_+}/m_{1_-}$.

\begin{figure}[!htb]
\begin{center}
\includegraphics[width=7cm]{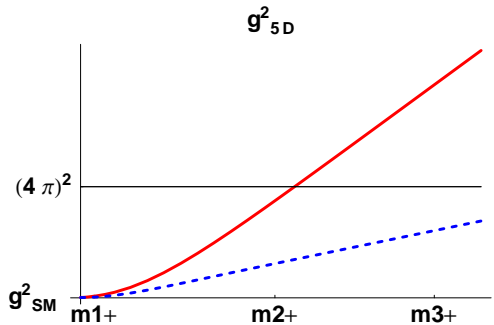}
\caption{\it The position dependent propagator smeared with $a^{-1}$ (solid red). It hits the strong coupling scale at the second heavy KK mass. For comparison, IR brane-to-brane propagators in the absence of IR BKT (dashed blue).}
\label{strongcoupling}
\end{center}
\end{figure}

Even though very large IR BKT's result in a sizable ratio between the first even and odd KK modes, which is
desirable from the phenomenological viewpoint, it would also imply the 5D gauge coupling $g_5$ is large
due to the relation $g_0=g_5/\sqrt{2r_{IR}}$, assuming that the zero-mode couples with the SM strength.
Therefore, if one demands the UV/IR hierarchy to be Planck-weak and/or the ratio $m_{1_+}/m_{1_-}$ to be sizable,
5D perturbativity may become an issue of concern. 
The strong coupling scale in the IR-UV-IR model can be estimated using the results from the single-slice AdS$_5$
setup by taking into account two facts: First, the physical volume in the IR-UV-IR model is twice as large, which
is reflected in the normalizations of the wave functions as well as the 
relation $g_5=\tilde{g}_5/\sqrt{2}$. Secondly, a single bulk field in the IR-UV-IR model contains
two towers of KK modes, both $(++)$ and $(-+)$ BCs, in the single-slice setup.
Consider an Euclidean propagator between two points $y_{1,2} \sim L$ in 4D momentum space.
It can be represented as $i g_{eff}^2(p^2)/p^2 $.   
At low energies, below the lightest KK mass we have $g_{eff}^2 \approx g_0^2$, but above the KK scale  the 
effective coupling grows with energy, which in the single-slice AdS$_5$ setup is \cite{Carena:2002dz} 
$g_{eff}^2 \approx e^{k L} \tilde{g}_5^2 p$ for one type of 
BC's.\footnote{Note
that, as seen in Fig.~\ref{profiles}, the regularly
spaced heavy KK modes (both odd and even) 
with mass $\sim \mkk$ tend to vanish 
at the IR brane 
due to the repulsion by very large BKT's. So,
if we consider the propagator between two points
localized  exactly on the IR brane, 
we will not find the above growth with energy
since the heavy modes do not contribute to
this propagator.
In order to include the effects of these heavier
KK modes giving the above growth of the
effective coupling with energy, we must consider 
the propagator with endpoints 
which are $\sim 1/ k$ 
(which is roughly the width of these KK profiles) away from the IR brane
thus accounting for the use of 
a smearing factor in Fig.~\ref{strongcoupling}.}  
In the IR-UV-IR setup the growth is twice as large
because both types of BC's are included. Therefore, defining the strong coupling scale $\Lambda$ by 
$g_{eff}^2(\Lambda) = 16 \pi^2$, one arrives at the estimate  
\beq
\Lambda \sim    e^{- k L} \frac{16 \pi^2}{g_5^{2}} \sim   \frac{8 \pi^2 }{ k L g_0^2} \mkk  
\left( \frac{ m_{1_-} }{ m_{1_+ } } \right)^2
\eeq
where we used $g_{0} \approx g_5/ \sqrt{2r_{IR}}$ and $m_{1_+}/m_{1_-} \approx \sqrt{ r_{IR}/L}$.
For example, setting $g_0^2 \sim 1/2$, $m_{1_-} /m_{1_+} \sim 1/2$ and  $k L\sim 30$,  we get $\Lambda \sim  \mkk \sim$
tens of TeVs. 
The strong coupling scale is far above the masses of the lightest even and odd KK modes, however it is not separated from the scale where the tower of evenly spaced KK modes sets in. 
These estimates are confirmed by the numerical analysis in Fig.~\ref{strongcoupling}. 
Thus there is no energy regime where the theory is effectively five-dimensional and weakly coupled (for that we would need $\Lambda \gg \mkk$).
As a compromise, we might need to lower the UV brane scale to some intermediate
scale (i.e., choose smaller $k L$) in which case
we loose the solution to the Planck-weak hierarchy
problem,
but we can still easily address the hierarchy between the weak scale and
(at least) the flavor scale $\sim 1000$ TeV.
\subsection{Fermions}
The Lagrangian for the fermions
\beq
{\cal L}_f =  \bar{ \psi }  \Gamma^M  \left( D_M - \eps(y) c k  \right) \psi
\eeq 
has the $Z_2$ symmetry $y \to - y$ with $\psi_{L,R} \to \gamma_5 \psi_{L,R}$. In the above
$\{\Gamma^M\}$ are the 5D Dirac matrices and $D_M$ is the
covariant derivative. As is familiar from the RS1 and UED setups, a bulk fermion mass term is odd under the reflection in $y \to -y$, therefore we need to include a bulk mass profile that is odd 
under $y \to -y$ and 
introduce the $c$ parameter such that $M_b=\eps(y) c k$. Notice however that in the conventional 
either flat or warped extra-dimensional setups, the physical domain is only from 0 to $L$ after the 
orbifold projection. So even though the bulk mass profile is odd under $y \to -y$, the mass term itself
is constant over the whole physical domain in $[0,L]$. In our case, the physical domain has been extended 
from $[0,L]$ to $[-L,L]$ and the mass profile would in fact include a jump at $y=0$. At this stage we will
not be concerned with the detailed origin of such a mass profile except to note that a plausible source
could arise from coupling to a scalar with a kink profile, similarly to the orbifold setup in Ref.~\cite{Kaplan:2001ga}.

As shown below, for the fermions we do not need BKTs to obtain a splitting between even and odd KK modes,  so we omit them in most of the following discussion.  
The IR boundary conditions require vanishing of one chiral component on the boundaries.   
Consider the case when the right-handed component vanishes: $\psi_R(L) = 0$; in this case there is a massless zero mode
for the left-handed component. The discussion for the case $\psi_L(L) = 0$ is in parallel, with $c \to -c$. 

Like the gauge field, a 5D fermion contains two KK towers with different UV boundary conditions:
\bea
\psi_L(x,y) &=& \sum_{n_+,n_-} a^{-3/2}f_{L,n_+}(|y|) \psi_{L,n_+}(x) + \eps(y) a^{-3/2} f_{L,n_-}(|y|) \psi_{L,n_-}(x)
\nn
\psi_R(x,y) &=& \sum_{n_+,n_-} \eps(y) a^{-3/2} f_{R,n_+}(|y|) \psi_{R,n_+}(x) + a^{-3/2} f_{R,n_-}(|y|) \psi_{R,n_-}(x)
\eea   
where the profiles satisfy the following coupled, first-order equations of motion
\begin{eqnarray}
\left(\partial_y +  \frac{a^\prime}{2a} + M_b \right) f_{L,n} &=& m_n a^{-1} f_{R,n} \\
\left(-\partial_y - \frac{a^\prime}{2a} + M_b \right) f_{R,n} &=& m_n a^{-1} f_{L,n}.
\end{eqnarray}
The massless zero mode $f_{L,0}(y)$ is even under reflection $y\to -y$. For massive
modes, the equations of motions imply that when the left-handed component has a symmetric profile 
under reflection, the 
corresponding right-handed chirality has an anti-symmetric profile, and vice versa. 
We insert an extra $(-1)$, in addition to the reflection in $y\to -y$, in the definition
of KK-parity for the right-handed chirality. In the language of the orbifolding, this extra minus sign 
could arise from performing the orbifold projection and is consistent with the definition of KK-parity in 
UED.

\begin{figure}[!htb]
\begin{center}
\includegraphics[width=7cm]{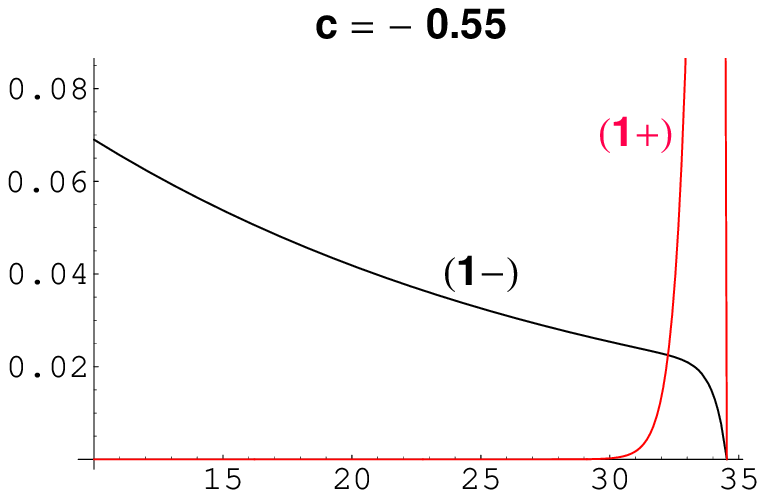}
\includegraphics[width=7cm]{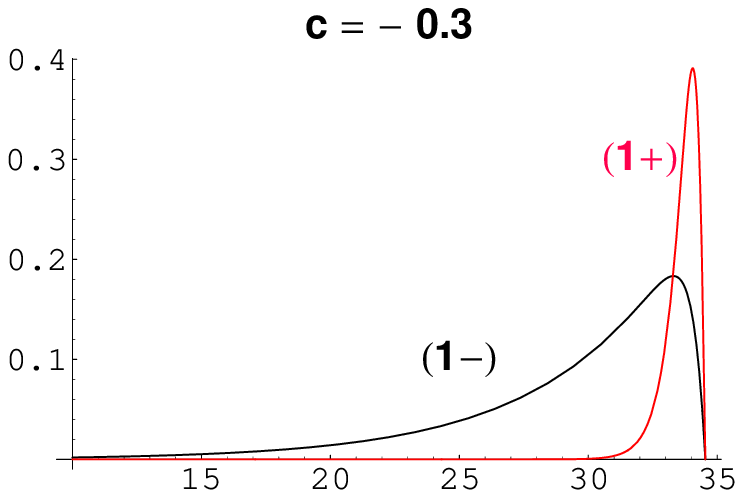}
\caption{\it Profiles of first odd (black) and even (red) KK fermions with RH chirality 
for two values of $c$.}
\label{Fermion_profiles}
\end{center}
\end{figure}

With the above definition of KK-parity, the {\em even} tower
has right-handed components that are anti-symmetric in $y\to -y$ and  
the ``continuity condition'' $f_{R,n_+}(0) = 0$, which can be interpreted as the boundary condition on the UV
brane.
The left-handed zero mode has the profile $f_{L,0} \approx e^{(1/2 - c) k y}$, which is localized towards UV for 
$c > 1/2$ and towards IR for $c < 1/2$.  
The massive KK-even modes start at $\sim \mkk$ for all values of $c$. 
For the {\em odd} tower
 the continuity condition reads $f_{L,n_-}(0) = 0$. 
The mass of the lightest odd state is 
\bea 
m_{1_-} &\sim& \frac{ \mkk }{\sqrt{ k L } } \; \hbox{to} \; \mkk  \qquad c \gtrsim -1/2
\nn 
m_{1_-} &\sim& \mkk e^{k L (1/2 + c)} \qquad c <  -1/2
\label{fermionodd}
\eea  
Thus, choosing $c <  -1/2$ we can generate a sizable splitting between the lightest even and odd KK modes without resorting to BKTs. In that case the RH profile is localized toward UV: 
$f_{R,1_-} \sim  e^{(1/2 + c) k y}$ (see Fig.~\ref{Fermion_profiles}).   
Note that the splitting can only be achieved if the corresponding zero mode fermion is sharply localized at the IR brane.
As is clear from the discussion, the zero mode fermion has $(++)$ BC's on the (UV, IR) branes. 
Changing its BC's from $(++)$ to $(-+)$ produces
a would-be zero mode that is very light, as the wave function is localized near the IR and insensitive to the BC at the UV,
which is nothing but the lightest odd mode. 
Typically, naturalness arguments require only the top quark KK modes  below $\simlt 1 \tev$  and the top quark is always 
localized toward IR, naturally giving light odd KK modes for it.
Hence, the even-odd splitting for 
KK fermions that we obtain by choosing $c$ appropriately
is sufficient for our purpose (this is different from the gauge case where the introduction of  large BKT's is necessary).  

For the light fermions there are two options: 
they can be localized near the UV or near the IR brane. 
The former setup allows us to simply address the Yukawa hierarchy and flavor issues, but, 
as we show below, is more constrained by EW data. 
\bi 
\item {\bf Light fermions near the UV brane.}

The light fermions can be localized near the UV brane by choosing the corresponding 5D mass parameter $c>1/2$. 
This yields naturally small couplings to the Higgs  localized in IR  and so the flavor hierarchy is addressed \cite{gn,gp}.   
At the same time, a severe flavor problem is avoided \cite{gp, Huber:2000ie}.
However, in this case, the coupling of  light 
fermions to the {\em lightest} even gauge KK mode is enhanced
compared to the SM gauge coupling. 
From  Eq.~(\ref{crat}), the coupling is given approximately by $g_{5}/ \sqrt{2L}$ which is enhanced with respect to 
the zero-mode coupling, $g_5/\sqrt{2r_{IR}}$, by a factor equal to the splitting between even and odd gauge KK's.
Integrating out the lightest even gauge boson will 
induce    4-fermion (flavor-preserving) operators  with the coefficient given by 
$\sim g_{5}^2 / (2L)\times  m_{ + }^{-2} \sim g_0^2 / m_{ - }^2$. 
Since the limit on the mass scale suppressing $4$-fermion operators is a few 
TeV \cite{Barbieri:1999tm},  the  EW data constrain the mass of the {\em odd} mode to be $\gtrsim$ a few of TeV.
Thus, with the light fermions  on the UV brane there is a tension between naturalness and electroweak precision data. 

\item{\bf Light fermions away from the UV brane.}

The alternative is to localize the light fermions away from the UV brane such that their
coupling to the lightest even gauge KK mode is suppressed. 
Such a localization for zero-mode fermions can be achieved either (i) 
in the   standard  way by choosing $c < 1/2$
or (ii) keeping $c \gtrsim 1/2$ and adding huge IR fermion kinetic terms\footnote{In this case, the fermionic profile is peaked towards the UV.
 However the dominant contribution to the normalization of the fermion zero-mode (and its coupling to gauge
modes) comes from the IR localized kinetic term.}  $k r_F > e^{(2c-1) k L}$. 
In the case of the light fermions localized very close to the IR brane, the coupling to the lightest even mode is 
smaller than the SM strength, see Eq.~(\ref{e.ireven}).
Consequently, constraints from four-fermion operators are not so stringent.   
In this case, the main constraint comes as usual from the S parameter 
and requires the lightest {\em even} mode to be heavier than a few TeV. 
In turn, the odd KK mode can still be lighter than a TeV in order to improve naturalness.

Nevertheless, with the light fermions localized in the IR, the flavor  hierarchy is not addressed in the usual fashion as in 
Refs.~\cite{gn,gp}. 
We also expect a severe flavor problem: the four-fermion flavor-violating operators from 
integrating out the cut-off physics are generically too large,
even though contributions from gauge KK exchange might be suppressed
due to the latter's repulsion from the IR brane where the light
fermions are localized.
Such large effects arise either from the cut-off suppressed operators in the bulk for the case (i), or are 
localized  on the IR brane in the case (ii).
To avoid flavor problems we should equip the model with additional flavor structures, see e.g. \cite{Rattazzi:2000hs}. 
\ei
\begin{figure}[!htb]
\begin{center}
\includegraphics[width=10cm]{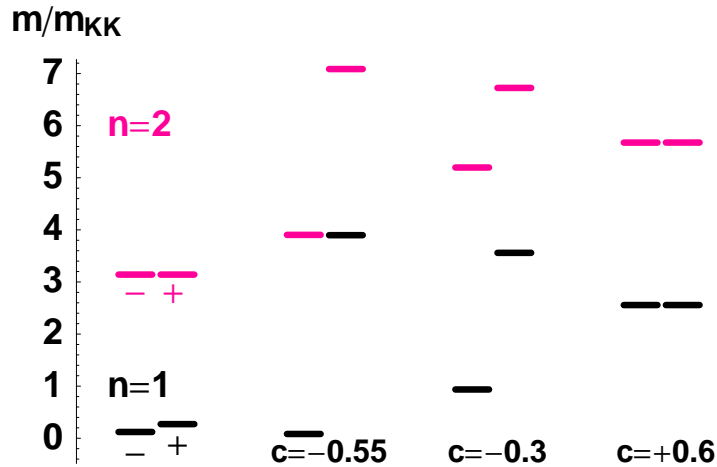}
\caption{\it KK mass spectrum. The first tower is for gauge bosons ($r_{IR}=4L$). The last three towers are for fermions with different $c$ parameters. The $n=1$ modes are black and the $n=2$ are pink. Each tower contains two sub-towers, the left one is for KK parity-odd modes, the right one for KK parity-even modes. }
\label{spectrum}
\end{center}
\end{figure}

\subsection{Dark Matter}
\label{subsection:dm}

KK parity implies that the lightest  KK-odd particle (LKP) is stable. 
There are two main possibilities: 
it could be either the lightest KK-odd gauge boson or the lightest KK-odd fermion (in the IR-UV-IR setup with large IR BKTs, the KK graviton is never the lightest mode).  
From our previous discussion and as illustrated in the spectrum of Fig.~\ref{spectrum}, the LKP can be a fermion if 
the $c$ parameter is  $c \lsim-1/2$, that is when the zero mode is sharply localized toward the IR brane.
From naturalness arguments, we expect the appearance of a light odd KK mode of the top quark. 
In particular, we expect that the only fermion having a  $c$-value close to $-1/2$ is the RH top quark.
However, in order to be a viable dark matter candidate, the LKP has to be electrically neutral and should interact weakly  and this discards the case where the lightest odd-KK top quark is the LKP.
The only possibility for fermionic LKP dark matter would be the KK partner of the RH neutrino, assuming the RH neutrino has the smallest $c \lsim -1/2$.   
This would mean the zero mode of the RH neutrino lives near the IR brane, which is not very well motivated since the neutrino is the lightest of the SM particles and we expect it to be localized in the UV. 
Therefore, in the following we do not consider the KK-odd fermion LKP case and we refer to 
\cite{Agashe:2004ci,Belanger:2007dx} for analysis of Dirac RH neutrino dark matter. 

Having concentrated on the lightest gauge boson as the LKP, there remain several options that lead to different interactions of the LKP.
Here we consider the situation in the KK parity symmetric version of the model of Ref.~\cite{Agashe:2003zs} where the electroweak symmetry is extended to  $SU(2)_L \times SU(2)_R \times U(1)_X$  and contains custodial symmetry. 
The model contains three neutral gauge bosons, $L_{1-}^3$, $R_{1-}^3$, $X_{1-}$, and the LKP could be a combination of those. 
In our setup with large brane kinetic terms, the masses of the lightest gauge states depend in the first place on the relative size of the IR BKTs $r_L$, $r_R$, $r_X$ for the three group factors. 
Unlike the minimal UED scenario, the one loop corrections to gauge boson masses play a secondary role (they are still relevant  though, because they split the masses of charged and neutral gauge bosons).  
Generically, the LKP will be embedded in the group factor with the largest BKT. 
The annihilation cross section of the LKP can be very different, depending whether the LKP is embedded in  $R_{1-}^3$ or $X_{1-}$, or whether it lives in $L_{1-}^3$. 

If the LKP is $X_{1-}$, it has no non-abelian gauge interactions whatsoever. 
If it is  $R_{1-}^3$ it does have non-abelian interactions, however vertices with the SM $W$ boson (who lives in $L_0^\pm$) are only induced by electroweak symmetry breaking and are very suppressed.  
Thus,  both of these cases are similar and, using the UED nomenclature, we refer to both as the KK photon LKP. 
In UED, the KK photon  annihilates dominantly into SM fermions with SM couplings \cite{Servant:2002aq} and its mass is predicted to be close to the 1 TeV scale to account for the observed dark matter abundance. 
In the model at hand, the situation is different due to different mass scales and non-trivial profiles along the extra dimension.
The lightest KK-odd gauge boson is peaked toward the IR brane
and it couples with the SM strength only to the SM fermions localized toward the IR brane.  
Furthermore, by $Z_2$ parity conservation, the interaction vertex with the light fermion must involve an odd  KK fermion.
The latter are typically very heavy in our setup, unless the corresponding SM fermion is sharply localized on the IR brane ($c < -1/2$).    
Thus, typically the LKP can annihilate efficiently only to top quarks.
For this reason, the annihilation cross section into fermions will be too small to support a TeV mass dark matter particle, unless all SM fermions are sharply localized toward the IR.

\begin{figure}[!htb]
\begin{center}
\includegraphics[height=3.cm,width=14cm]{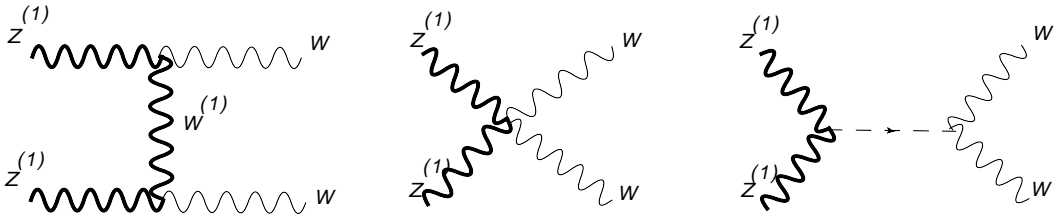}
\caption{\it Diagrams contributing to the annihilation of the KK $Z$. 
In the first diagram  both $t$ and $u$ channels should be included. And in the case of the $SO(4)$ model, both vector $V^{\pm}$ and axial $A^{\pm}$ charged gauge bosons are exchanged.}
\label{Feynmanndiagrams}
\end{center}
\end{figure}

The possibility that the LKP is $L_{1-}^3$, which we refer to as the KK $Z$, appears more promising.  
In UED, KK $Z$ is usually not considered as the LKP. The reason  is that, without BKTs, the KK photon is lighter than KK $Z$ due to one-loop corrections to KK masses  \cite{Cheng:2002iz}.
In the present setup, however, there is no reason to reject  the KK $Z$ scenario. 
The most important point is that the KK $Z$ has non-abelian gauge interactions with the SM $W$ bosons.  
More precisely, we have the trilinear vertex: 
\beq
\label{e.gtv} 
{\cal L}_{3} \approx - i g_L   
(\pa_\mu L_{1-,\nu}^3 - \pa_\nu L_{1-,\mu}^3)  L_{1-,\mu}^+ W_\nu^-   + \dots  
\eeq   
and the coupling here is the SM $SU(2)_L$ coupling.  We also have the quartic vertex: 
\beq
\label{e.gqv}
{\cal L}_{4} \approx 
-  g_L^2  L_{1-,\mu}^3 L_{1-,\mu}^3 W_\nu^+ W_\nu^-  + \dots  
\eeq 
In the above, we neglected the effects of electroweak symmetry breaking. 
These couplings lead to the annihilation diagrams shown in Fig.~\ref{Feynmanndiagrams} and the annihilation cross section into $W^+W^-$ is \cite{Burnell:2005hm}
\beq
\sigma_{L^3 L^3 \rightarrow W^+ W^- }
\ = \ \frac{g_L^4}{18\pi m^2 s^3 \beta^2} \left[ - 12m^4 (s- 2m^2 )L + s \beta (12m^4+3s m^2+4s^2) \right] 
\eeq 
where $\beta^2 = 1 - 4 s^2/m^2$, $L = \log[(1 + \beta)/(1-\beta)]$.  
The KK $Z$ couples also to the Higgs boson which is localized on the IR brane. 
According to Fig.~\ref{profiles}, the lightest odd gauge boson couples with the same strength as the zero mode to the IR brane.
Thus, the coupling to the Higgs has the SM strength. 
Annihilation via the Higgs boson yields only a small correction  (however, the coupling to the Higgs will be relevant for direct detection). 
For the same reasons as in the KK photon case, we do not expect the cross section for annihilation into fermions to be sizable.   Finally, annihilation into $ZZ$ and $hh$ are comparatively negligible. 

An interesting variation of the KK $Z$ LKP is the case when the gauge 
couplings and the BKTs for $SU(2)_L$ and $SU(2)_R$ are equal, which may occur 
if the model displays $SO(4)$ symmetry (that may be a consequence of the 
larger underlying $SO(5)$ symmetry as in \cite{Agashe:2004rs}). In the $SO(4)$ 
invariant case, $L^3_{1-}$ and $R^3_{1-}$ are degenerate in the limit of no EW 
breaking. Electroweak breaking lifts the degeneracy, and it picks up the 
vector combination $V^3 = L^3_{1-} + R^3_{1-}$ as the LKP, while the axial 
combination $A^3 = L^3_{1-} - R^3_{1-}$ which couples to the Higgs on the IR 
brane is heavier. The couplings of the LKP to the SM $W$ bosons are reduced by 
one half with respect to the previous case. Moreover, the annihilation via 
$t$- and $u$-channel exchange of the charged axial gauge bosons should be 
taken into account. All in all, the annihilation cross section due to 
non-abelian gauge interactions is reduced by one quarter,  \beq \sigma_{ V^3 
V^3 \rightarrow W^+ W^- } = {1 \over 4} \sigma_{ L^3 L^3 \rightarrow W^+ W^- } 
\eeq    Furthermore, the vector LKP does not couple to the Higgs boson at all. 

\begin{figure}[!htb]
\begin{center}
\includegraphics[height=7.cm,width=11cm]{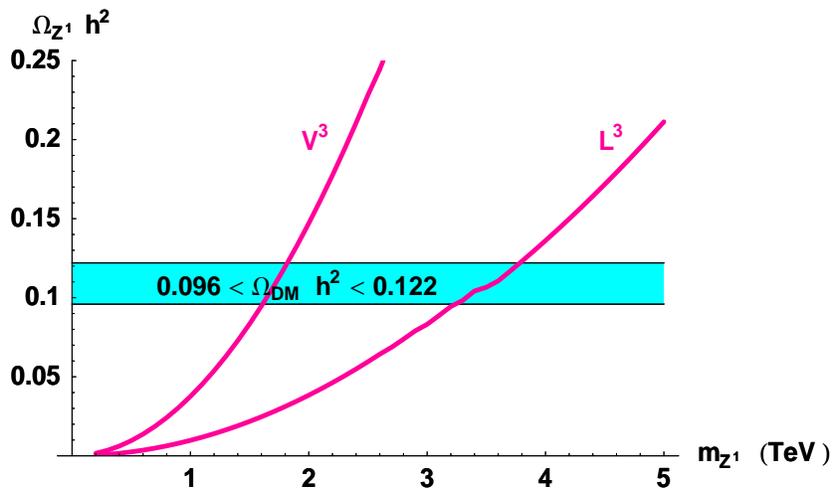}
\caption{\it Relic density prediction as a function of the $Z^1$ mass in two cases. 1) $Z^1$ is $L^3_{1-}$ and has SM couplings. 2) $Z^1$ is $V^3=L^3_{1-}+R^3_{1-}$ with $g_L=g_R$. Only self-annihilation into $W^+W^-$ is included.}
\label{relic}
\end{center}
\end{figure}

Like in UED, we assume that the reheat temperature is at least a few tens of GeV so that the relic dark matter 
abundance  follows from the standard thermal freeze-out procedure and is entirely determined by the annihilation 
cross section of the LKP. 
In the generic KK $Z$ scenario, this leads to $m_{LKP}\sim 3.5$ TeV to obtain  the correct relic abundance 
(see Fig.~\ref{relic}).
This mass scale is quite high and would signify that the little hierarchy problem is not solved in our model. 
The situation is better in the $SO(4)$ invariant case, where the reduction of the annihilation cross section leads 
to the smaller LKP mass,  $m_{LKP}\sim 1.7$ TeV.  
Moreover, this mass scale could be further reduced if co-annihilation is taken 
into account \cite{Servant:2002aq}. We indeed expect the next lightest KK 
modes (NLKP) $W^{\pm}_{1-}$ (as well as $A^3_{1-}$ and $A^{\pm}_{1-}$ in the 
$SO(4)$ model) to be close in mass to the LKP. The relevant self-annihilation 
cross sections of the nearly degenerate states as well as the co-annihilation 
cross sections were computed in  Ref.~\cite{Burnell:2005hm} to study 
co-annihilation effects in KK photon dark matter but were not used to study KK 
$Z$ dark matter. This issue is, however, model-dependent and here we do not go 
beyond  the rough estimate obtained without co-annihilation.  

\begin{figure}[!htb]
\begin{center}
\includegraphics[height=6.cm,width=10cm]{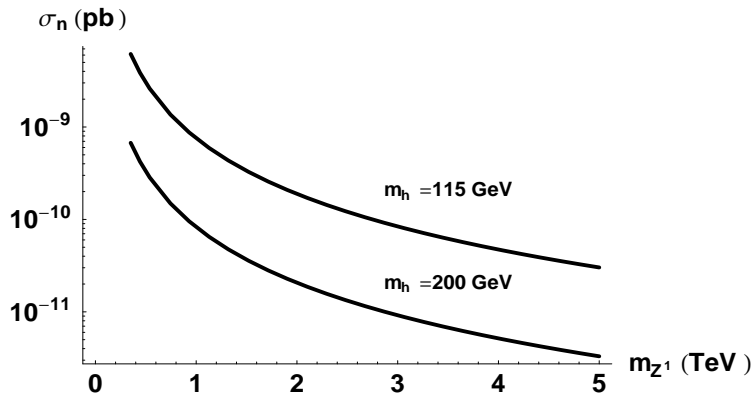}
\caption{\it Spin-independent elastic scattering of $Z^1$ ($=L^3_{1-}$) on nucleon.}
\label{direct_detection}
\end{center}
\end{figure}

Direct detection of KK $Z$  from its elastic scattering off a nucleus in underground detectors such as CDMS or 
XENON will be very challenging, and in the $SO(4)$ model, it is hopeless since in this case there is no coupling 
to the Higgs.   
To predict the rates for direct detection of a heavy $Z^1=L^3_{1-}$, we can use the same analysis as the one for 
UED \cite{Cheng:2002ej,Servant:2002hb}, replacing  the hypercharge coupling by the $SU(2)_L$ coupling. 
 In addition, we can remove the effects from fermion interactions and only take into account the elastic scattering 
 from $t$-channel Higgs exchange. The spin-independent elastic scattering cross section on a nucleon is
 \begin{equation}
 \sigma_n=\frac{m_N^2}{4\pi(m_{Z^1}+m_N)^2}\left[Zf_p^{Z^1}+(A-Z)f_n^{Z^1}\right]^2 \frac{m_{p,n}}{A^2\mu} \ \ \mbox{where} \ \  f^{Z^1}_{p,n}=m_{p,n}{\sum} f^{p,n}_{T_q}\frac{g^2}{2m_H^2} 
 \end{equation}
where  $A$ and $Z$ are the number of nucleons and protons in the nucleus, $m_{n,p}$ is the mass of the proton or 
neutron, $\mu=m_N m_{Z^1}/(m_N+m_{Z^1})\sim m_N$ is the reduced mass of the WIMP-nucleus system and $f^{p,n}_{T_q}$ 
are the usual nucleonic matrix elements.
We therefore have a $(g/g')^2$ enhancement  compared to UED but also a suppression from the higher mass. So, at the 
end, the predictions are of the same order as the ones from UED, where elastic scattering  is also dominated by Higgs 
exchange, unless some enhancement effect takes place from KK fermion exchange if we force  a mass degeneracy 
between $\gamma^1$ and KK quarks. As shown in Fig.~\ref{direct_detection}, for $Z^1$ masses of order 3--4 TeV, 
the nucleon-$Z^1$ spin-independent cross section is smaller than $10^{-10}$ pb, well below  
the projected sensitivities of near-future experiments.

\subsection{Collider signatures}

As suggested in the previous subsection, an LKP mass at  1 TeV is possible if the $SU(2)_R$ component of the LKP is increased, which makes its effective coupling to the SM smaller and its relic density compatible with observations for a mass smaller  than in the case of a pure $SU(2)_L$ coupling.
We have also argued that if $Z^1$ is indeed the LKP (either $L^3_{1-}$ or $V^3_{1-}$), we expect the next lightest KK modes (NLKPs) $W^{\pm}_{1-}$ (as well as $A^3_{1-}$ and $A^{\pm}_{1-}$ in the $SO(4)$ model) to be close in mass to the LKP.
There is, on the other hand, a large mass splitting between these modes  and the other KK states (even gauge KK modes and KK fermions other than the KK top) --unless the fermions are localized on the IR brane. This follows from our prejudice that, as required by EW precision tests, only these  gauge fields have large brane kinetic terms. Therefore,
 we expect that only the KK top, the  LKP and nearly degenerate gauge KK modes to be produced at LHC. This is quite different from the usual UED phenomenology where masses of all first level KK modes are of the same order. The UED implications for collider phenomenology  were discussed in  \cite{Cheng:2002ab}. Pair production of KK fermions lead to cascade decays and final states with leptons, jets and missing energy, very much like supersymmetric signatures.
The distinction in our setup  is that the only SM particle the LKP couples significantly to is a $W$ so that we always end up with at least one $W$ in the final state.
 Pair production of $t^1_R$  leads to $t\overline{t}Z^1Z^1$. This eventually leads to jets, leptons and large missing energy like in SUSY and UED but one way to probe this LKP scenario would be to  reconstruct $W$ and $t$ candidates. 

\section{UV-IR-UV Model}
\label{UVIRUV}

In this section we consider another setup with $Z_2$ parity where we glue the two AdS$_5$ slices 
in the IR region (instead of the UV region as considered before).
We call this setup the UV-IR-UV model. The metric is 
\begin{eqnarray}
( ds )^2 = ( dy )^2 + e^{ + 2 k \left( | y | - L \right) } ( d x )^2,
\end{eqnarray}
The warp factor has a {\em minimum} at the midpoint, which is now referred to as the IR brane, 
while the two end-point branes at $y = \pm L$ are UV branes. 

The above metric is a solution of the 5D Einstein's equations with a negative bulk cosmological constant, 
once the two UV branes have equal positive tensions while the IR brane has a negative tension. 
The problem is that  the radion is a ghost due to the negative tension on the IR brane 
(in the original Randall-Sundrum setup the would-be ghost is projected out by the boundary conditions on the negative tension brane). 
One might try to avoid the instability problem by adding a large graviton kinetic term on the IR brane that would 
give the radion a large enough right-sign kinetic term. A large 4D kinetic term for the graviton 
is reminiscent of the DGP model in 
Ref.~\cite{Dvali:2000hr}. However, it is known that the DGP models may still have a ghost in the gravity 
sector \cite{Luty:2003vm}.
An alternative is to consider a continuous metric in which case there is no 
need for a negative tension brane.\footnote{ In fact, it is also possible to 
find a continuous warp factor qualitatively similar to the IR-UV-IR setup 
without the UV brane in the middle, which has the behavior of $1/\cosh(2ky)$; 
see Ref.~\cite{Low:2000pq}.} For example, the  ``cosh'' metric $( d s )^2 = ( 
d y )^2 + \cosh ( 2 k y ) /\cosh (2 k L) ( d x )^2$ yields a spectrum that is 
qualitatively similar to that of the UV-IR-UV model. The cosh metric is a 
solution to the 5D Einstein's equations  in the presence of a negative $T_{ 55 
}$ in the bulk  (and two positive tension branes as 
before)\cite{Kanti:2000rd}. A possible source of $T_{ 55 } < 0$ was proposed 
in Ref.~\cite{Mukohyama:2000wq} using a conformal scalar. However, it was 
claimed that in this model the radion is a {\em tachyon} 
\cite{Hofmann:2000cj}, instead of a ghost. One might be tempted to invoke the 
usual mechanisms such as Goldberger-Wise or Garriga-Pomarol 
\cite{Goldberger:1999uk} to stabilize the  radion, i.e., make its (mass)$^2$ 
positive. However, the worry is that a back-reaction on the metric is so large 
that the warp factor may lose the qualitative UV-IR-UV behavior: whatever 
lifts the tachyonic mass of the radion would also makes a non-negligible 
contribution to the stress-energy tensor such that in the end $T_{55}$ becomes 
positive again.

In fact, one can prove a $c$-theorem on the behavior of the warp factor on very general ground. If we write the warp
factor as $a(y) = e^{2A(y)}$, using 5D Einstein's equations one can show that the weak energy condition implies
\cite{Freedman:1999gp}:
\begin{equation}
A''(y) \le 0.
\end{equation}
Clearly the IR-UV-IR setup, in which $A'(-\infty)=2k$ and $A'(\infty)=-2k$, satisfies this $c$-theorem, whereas the UV-IR-UV
setup violates the $c$-theorem. In other words, negative energy sources violating the weak energy condition must be present in order for the UV-IR-UV setup to be 
a solution of the Einstein's equations. In Ref.~\cite{Mukohyama:2000wq} such a negative energy source is provided by the
casimir energy. Obviously there are other examples of negative energy sources in nature such as the dark energy driving the
expansion of our universe. It remains to be seen if one could find a model with the UV-IR-UV-like warp factor that is free from the ghost or the tachyon. 

In the following we very briefly explore the phenomenological features of the UV-IR-UV setup, since it is an obvious alternative to the IR-UV-IR setup, while keeping in mind
that we are not aware of any satisfactory solutions to the issue of instability.
Based on the previous discussion, it is clear that the 
spectrum of the UV-IR-UV model contains the even modes $(++)$ and the odd modes $(+-)$
(note that these BC's refer to a {\em single} slice of AdS$_5$).
One can find that the lightest even gauge KK mode mass is $m_{1+} \sim m_{ KK }$,
whereas the lightest odd gauge KK mode mass is  $m_{1-} \sim m_{ KK }/\sqrt{k L}$
(as mentioned earlier).
If the 5D model addresses the Planck-weak hierarchy,  
a large  splitting between even and odd gauge KK modes is automatic;
there is no need for large BKTs in this setup. This would have been a very desirable feature phenomenologically.

A peculiar feature of this model is the presence of a very light massive graviton state 
(in addition to the massless graviton). 
The mass of the lightest odd mode of the graviton 
turns out to be $m_{1-} \approx 2 \sqrt {2} e^{- k L} \mkk$. 
Thus, it is suppressed with respect to the KK scale by the factor equal to the UV-IR brane hierarchy.
If this hierarchy is Planck-weak, the lightest KK graviton mass is of order $10^{-3}$ eV.   
This small mass comes because the  would-be zero mode  graviton from the $(+-)$ sector  is highly localized near the UV brane, with the wavefunction near the IR brane suppressed by  $e^{- k L}$ (just like the actual zero-mode from the $(++)$ sector).
For this reason it is insensitive to changing the BC on the IR brane. 
Equivalently, we could think of the mass as resulting from adding a longitudinal graviton mode near the IR brane to lift the would-be zero-mode and the small overlap between the transverse and longitudinal modes gives a small mass.
 There are many worries over such an exponentially light massive graviton, which all result from the
vDVZ discontinuity \cite{van Dam:1970vg}. Historically a tiny mass for the graviton has been ruled out by bending of light
around the sun. However, the assumption there was based on the classical one graviton exchange between the sun and the
photon. In the particular setup we are considering here, such an interaction is forbidden by the KK-parity because
the light graviton is an odd mode under KK-parity, and therefore the experimental constraint might be loosened. Even if
one were able to get away with the constraint from bending of light, a very light massive graviton has been shown to be 
plagued by the strong coupling issue. For a graviton with mass $m_g$, it is shown in Ref.~\cite{ArkaniHamed:2002sp} that
the highest energy scale one can delay the strong coupling problem to is $\Lambda_3=(m_g^2 M_{pl})^{1/3}$, 
where $M_{pl}$ is the 4D Planck scale. For a massive graviton at $10^{-3}$ eV, this would translate into $\Lambda_3 \approx$
1 GeV, at which scale we lose control of the gravity sector.

To summarize, even though the UV-IR-UV setup provides a large splitting between the first even and odd KK gauge bosons, 
which is a nice phenomenological feature, the gravity sector seems to suffer from various instability and strong coupling 
problems.

\section{Conclusion}
\label{conclude}

In this work, we considered the possibility of implementing  Kaluza--Klein parity in a warped geometry. The point is that
KK-parity can allow for a lower mass scale for the new particles while
satisfying the electroweak constraints. Besides, collider signatures of the resulting models are different from 
either of the two popular extra-dimensional models: the UED and the RS models. In UED, there is KK-parity and KK number 
conservation so that
the mass scale of new particles can be as low as 300 GeV \cite{Appelquist:2000nn}
and the LKP can be a good dark matter candidate \cite{Servant:2002hb}. 
Moreover, because of the flat geometry, the KK mass spectrum is evenly spaced and KK parity imposes 
pair-production of KK-odd particles. Despite the nice feature of allowing for new particles at masses as low
as several hundreds GeV, UED models do not seem to address any hierarchy problems.

On the other hand, in the RS setup, where both electroweak symmetry breaking and the Planck-weak hierarchy are addressed, there is no $Z_2$ parity and 
all new particles can be produced singly. 
However, precision electroweak  and flavor tests constrain the mass scale of KK gauge bosons to be heavier than 2 - 3 TeV (KK fermions are allowed to be lighter, in some circumstances). 
Furthermore, the first few KK masses are not evenly spaced due to the  warped background. 
Finally, there is no stable KK state unless an extra non-geometrical symmetry is imposed\footnote{There could be a KK dark matter candidate in RS models, following from a $Z_3$ symmetry imposed to solve the proton decay problem \cite{Agashe:2004ci}, but this is not a symmetry of geometrical origin unlike in UED.}. 

The  ``warped KK-parity''  setup we considered is the hybrid
of the two scenarios: KK parity allows for a light KK mode compatible with electroweak precision tests.  KK-odd particles need to be pair-produced, and the first few KK masses are not evenly spaced.

All Standard Model extensions which possess a new conserved quantum number at the TeV scale
share very similar collider phenomenology  \cite{Cheng:2003ju, Cheng:2002ab}. 
Pair-production of new (colored) particles lead to multiple jets $(\ge 2)$ and 
missing  energy signals from the dark matter candidate as well as isolated leptons  from cascade decays. In contrast, in models
without a new symmetry, not every event involving production of new particles would be associated with multiple jets and missing energy.
From this perspective, it is natural to wonder whether phenomenologies of models with warped extra dimension would always
fall into the category of single-production of new particles, 
in which case observations of only events with a large 
multiplicity of jets and missing energy would automatically disfavor warped extra-dimension, 
or there exists variants which would again always
produce events with multiple jets plus missing energy. 

In this work we made a first attempt toward studying the above question. Ideally, we would like to implement the good features of UED, namely KK modes below a TeV
and dark matter, in a warped background (so that the Planck-weak hierarchy is addressed) without 
giving up some of the virtues of warped extra-dimension such as fermion and Higgs localization. 
The first point to address is that, for a single slice of AdS$_5$, the warp factor is clearly 
not symmetric under reflection
about the midpoint of the extra dimension.
Therefore,
\bi

\item
we glue two physically distinct slices of AdS$_5$ and impose the symmetry 
interchanging the two AdS$_5$ slices.

\ei
In such a construction, the mass eigenstates can be divided into two classes with different symmetry properties. 
For any given level $n$ in the KK decomposition, there are
KK-even modes ($n+$), whose profiles are symmetric under reflection around the mid-point of the extra dimension,
and  KK-odd modes ($n-$) with anti-symmetric profiles.
KK-odd modes can only couple in pairs to the KK-even modes
and the low-energy, four-dimensional effective theory has  KK parity. 
It is important to stress that our construction does not implement  
approximate KK number  conservation, 
which would require that the fermion zero-modes and Higgs vev have a flat profile
in the warped extra dimension. 
However, we cannot give up on the localization of the Higgs 
profile near the IR brane if we want to solve the Planck-weak hierarchy 
problem so KK number conservation is definitively lost in our approach. 
Therefore,  while the odd modes are allowed to be lighter than a TeV, we need 
the even modes to  be heavier than a few TeV since KK parity by itself is not 
enough to satisfy EW precision tests. To achieve that, we have to impose 
further requirements on our setup. Namely,  
\bi

\item
we need  to obtain a sizable hierarchy, at least a factor
of a few,  between the lightest KK-even mode and the lightest KK-odd mode.
\ei

There are two distinct ways to realize our idea, depending on whether the two slices are glued 
at the UV or IR brane, leading to the IR-UV-IR and the UV-IR-UV models:
\bi

\item
In the IR-UV-IR model 
the splitting between even and odd gauge KK modes can only come from very large IR brane kinetic terms. 
The dark matter particle can be identified with the lightest KK partner of the $Z$ boson (the KK photon would not lead to the correct abundance since its couplings to the SM are different from the UED case) and the predicted relic abundance is in the correct range. However, there are two problems with this setup. 
One is that large IR brane kinetic terms create a certain tension with perturbativity and the regime where 
the 5D theory remains weakly coupled is rather narrow. 
The other problem is that for light fermions localized close to the UV brane the constraints from 
electroweak precision tests are still quite severe. The EW constraints can be softened by localizing fermions close to 
the IR brane, but then the flavor problem cannot be addressed by utilizing the different localizations of fermions
along the extra-dimension. Additional flavor symmetries need to be implemented, which we do not discuss in the present
work. 

\item
An apparent alternative to the IR-UV-IR setup is to glue the two slices of AdS$_5$ together in the IR region instead.
In such a UV-IR-UV model  the desired splitting between even and odd gauge is naturally obtained without brane kinetic terms. 
However, when gravity is included the radion becomes a ghost and it is a challenge to make the UV-IR-UV setup stable 
gravitationally. A related issue is the appearance of a very light massive graviton, which poses a very low strong
coupling scale around 1 GeV, at which the gravity sector already needs UV completion.

\ei

Thus, the IR-UV-IR setup seems a more promising approach to incorporate KK-parity in warped
extra-dimensional models, even though to obtain a sizable splitting between the lightest KK-odd and -even modes 
and avoid the strong coupling problem in 5D at the same time, one may need to move the UV brane to an intermediate 
scale below the Planck scale. This may be a drawback compared with the traditional RS models, but certainly 
is an improvement over the UED in which the hierarchy problem is simply not addressed at all. In addition,
the model discussed so far still requires additional mechanisms to address the issue of flavor violation. 
We proceeded in an exploratory spirit, focusing mostly on highlighting
the important issues or challenges in model-building, and hope to present a tool-kit 
for model-building along these lines. Moreover, we adopted a
phenomenological approach without concerning ourselves with whether the new particles stabilize the Higgs mass by canceling
the quadratic-divergent contributions from the standard model particles. 
However, in the appendix, we show in toy models how divergences in the SM Higgs
mass can be canceled by the lightest odd mode thus, possibly, providing a solution to the little hierarchy problem.
It certainly will be interesting to look into more details of how the 
requirement of Higgs mass cancelation would affect various constraints and 
phenomenology of the setup.

There are several non-supersymmetric approaches on how new particles could stabilize the Higgs mass,  which are all related directly or
indirectly (via the AdS/CFT conjecture) to models with warped extra dimension. Some of the more popular ones are the
gauge-Higgs unification \cite{Hosotani:1983xw}, the holographic Higgs models \cite{Contino:2003ve}, and the little Higgs
theories \cite{ArkaniHamed:2001nc}. Even though a $Z_2$ parity, the $T$ parity, has been implemented in the little Higgs
theories \cite{Cheng:2003ju}, no such attempts have been made with regard to the first two classes of models. Clearly, our work
could be viewed as an initial step toward that direction. In addition, it also seems likely that the IR-UV-IR setup could
serve as a possible UV completion of the little Higgs theories with T parity, without resorting to supersymmetrized linear
sigma models above 10 TeV. Much work remains to be completed.

\section*{Acknowledgments}

We thank Nima Arkani-Hamed, Hsin-Chia Cheng, Takemichi Okui, Riccardo Rattazzi, Tim Tait,
John Terning, Raman Sundrum
and Carlos Wagner for discussions. I.~L. acknowledges hospitality of the CERN
Theory Group during his visits in which part of this work was initiated and completed.
K.~A. was supported in part by the U.~S.~DOE under
Contract no. DE-FG-02-85ER 40231. I.~L. was supported in part by U.~S.~DOE under contract
DE-AC02-06CH11357 and by National Science Foundation under grant PHY-0653656.
A.~F. was partially supported by the EC contract MRTN-CT-2004-503369 for the years 2004-2008 and the MEiN grant  1 P03B 099 29 for the years 2005-2007. 

\vspace{1cm}

\appendix

\section*{Appendix A: CFT Interpretation}
\label{CFT}

We discuss here the CFT interpretation of the 5D models with KK parity considered in this paper, focussing only 
on the IR-UV-IR model since the UV-IR-UV model suffers from the instability in the gravity sector. 
In that model, we have two towers of the gauge fields: the even $(+ -)$ and the odd $(- -)$, where 
the Dirichlet boundary condition in the IR is effectively due to a large BKT. 
In addition, there is an even zero mode effectively localized near the IR (in the sense that its normalization is dominated by the IR BKT
even though it has a flat profile) and a light odd mode
localized near the IR brane. 

Gauge symmetry in the 5D bulk is dual to a global symmetry of a 4D CFT \cite{Witten:1998qj}, 
with $(-)$ on the IR brane dual to a spontaneous breaking of the global symmetry at
the TeV scale \cite{Contino:2003ve}. 
On the 4D side, the spontaneous breaking results in the presence of a Goldstone boson which is 
a composite of the CFT. 
Consider first the  $(+ -)$ tower in isolation. 
The $(+)$ BC on the UV brane is dual to a gauging of the global symmetry
with an external gauge field  \cite{Arkani-Hamed:2000ds}.
This external gauge field eats the Goldstone boson and becomes massive. 
The mass scale is set by the coupling of the external gauge field 
to the CFT in the IR,
which is given by $g_{ ext. } \sim g_{ \rho } / 
\sqrt{ \log(M_{PL}/\tev) }$, where $g_{ \rho }$ is the coupling
of the heavy spin-1 composites or ``$\rho$ mesons''
(assuming that the low-energy external gauge coupling is 
dominated by IR-free running due to the CFT sector).
The coupling $g_{ ext.}$ is dual to $\tilde{g}_5/\sqrt{L}$.\footnote{We remind the reader that $\tilde{g}_5$ is
defined as the 5D gauge
coupling in the single-slice AdS$_5$, see Eq.~(\ref{new5daction}).}
On the other hand,  the masses of 
heavy
spin-1 composites, which are dual to the typical heavy gauge KK modes, is set by 
$g_{ \rho }$,  which is dual to $\tilde{g}_{5} \sqrt{k}$.
Consequently, the mass splitting between the first and second KK mode is enhanced
by the factor $g_{ \rho } / g_{ ext. } \sim \sqrt{ \log( M_{PL}/\tev ) }$ 
(the logarithm of the large hierarchy) as we found in the
5D calculation.\footnote{ A similar explanation
can be given for the lightness of $(-+)$ fermion mode for
$c \lesssim -1/2$. First, note that
$(-+)$ spectrum is same as $(+-)$ spectrum
with {\em opposite} value of $c$, i.e., $c \gtrsim 1/2$
(see, for example, \cite{gp}).
Then, the CFT interpretation
is similar to that  of gauge field with $(+-)$ BC: 
an external fermion marries a composite fermion and it can be shown that 
the external fermion is even more weakly coupled to the CFT than in the case of the gauge field, resulting in an ultra-light mode from the marriage of
external with composite fermion \cite{Contino:2004vy}.}

The even zero-mode, being effectively localized in the IR, is dual to a massless CFT {\em composite} 
gauge field (like in the original RS1).
This gauge field also couples to the Goldstone boson, with the coupling 
$g_{ comp. }$, different from $g_{ \rho }$ and dual to 
$g_{IR} = \tilde{g}_5/\sqrt{r_{IR}}$.
For 
$g_{ ext. } \gg g_{ comp. }$, corresponding to the  limit of very large BKT's we are focusing on,
$g_{ IR } \ll \tilde{g}_5/\sqrt{L}$,
it is the external gauge field which marries the Goldstone boson, leaving the composite gauge field massless 
and vice versa for $g_{ comp. } \gg g_{ ext. }$.
The interpretation of the $(--)$ tower in isolation  is similar to that of the even tower above, 
except that the composite Goldstone boson always marries the composite gauge field since there is no external 
one in this case -- the absence of external gauge field is dual to the $(-)$ BC on the UV brane.

Finally, when we combine the two towers, the 4D dual interpretation
is that there are {\em two} identical CFT's (dual to the two
AdS slices related by KK parity).
Each CFT breaks a global symmetry 
spontaneously in the IR, resulting in one Goldstone boson from each
CFT and each CFT also produces a massless composite gauge field. 
A {\em single} external gauge field couples to {\em both}  CFT's
and, in the limit of very large BKT's that we are interested in, it
marries one linear combination of the Goldstones to become the massive $1_+$ mode.
One combination of the two composite gauge fields marries the other combination of the Goldstones to become the  
massive $1_-$ mode. 
The other combination of the {\em composite} gauge fields remains massless and is dual to the zero-mode gauge field.

\section*{Appendix B: Solution to the Little Hierarchy Problem}

In this appendix, we discuss how our ``warped KK-parity'' setup could 
address the little hierarchy problem such that the SM Higgs mass can be cut-off by
the lightest odd KK mode of the gauge field. At first we sketch a general argument based on 
the low-energy effective theory, i.e. the three-site model mentioned in Section 3, and
then support the intuition from the three-site model by an explicit computation in a 5D toy setup,
in which the Higgs originates
as the fifth component of a 5D gauge field ($A_5$).

\subsection*{B.1  Low-energy Perspective}
We begin by stressing that a new $Z_2$ parity at the TeV scale does not 
interfere with cancelations of quadratic divergences in the Higgs mass by the 
new particles, no matter whether the new particles are even or odd under the 
new parity. 
As explained in Ref.~\cite{Cheng:2003ju}, loop diagrams involving interaction vertices with two new particles are sufficient to engineer cancellations of  the quadratic divergences. 
Such interactions are always  allowed by the $Z_2$ parity as long as the two new particles involved are both 
odd (or even) under the new parity. Therefore it should be clear that whether 
the lightest gauge KK mode can stabilize the Higgs mass is entirely a question 
of engineering the cancelation through mechanisms such as the ``collective 
breaking'' in the little Higgs theories, or equivalently non-locality in the 
extra-dimension in Higgs as the $A_5$ theories, and as such is independent of 
the charge of the gauge KK mode under the new parity.

On the other hand, it is a legitimate question to ask if one can be sure that the Higgs mass is always 
cutoff by the {\em lightest} gauge KK mode. This is a question that goes into the heart of employing
all the different mechanisms to stabilize the Higgs mass, for if Higgs mass is cut off only by the 
second or higher gauge KK mode, the little hierarchy problem would not be solved at all. There
have been many studies on such a question; see for example Ref.~\cite{ArkaniHamed:2001nc}. 
At the risk of repeating what many experts have already 
known, we try to adapt the arguments to our particular setup of warped KK-parity.

Since we are interested in a ``low-energy'' question, in that whether the 
Higgs mass is cut off by the lightest gauge KK mode, it suffices to consider a 
``low-energy'' effective theory by using a three-site moose model as discussed 
in Section 3. In fact, Ref.~\cite{ArkaniHamed:2001nc} discussed a $N$-site 
moose model with uniform gauge couplings and decay constants, which can be 
considered as deconstructing a flat extra-dimension. There the authors 
computed the Coleman-Weinberg potential of the scalar corresponding to the 
zero mode of $A_5$ and showed explicitly that its mass is cut off by the 
lightest massive vector boson. Coincidentally, because of the flatness, there 
is KK-parity defined as the reflection with respect to the midpoint of the 
moose diagram, and the lightest massive vector boson is odd under the 
KK-parity. This supports the argument that KK-parity and stabilization of 
Higgs mass do not interfere with each other and a KK-odd massive gauge boson 
{\em can} in fact cut off the quadratic divergence in the Higgs mass.

In the context of our setup, we consider an extra-dimensional toy model with 
the $SU(2)$ bulk gauge symmetry broken down to $U(1)$ on the boundaries. The 
low-energy effective theory of such a toy model corresponds to a three-site 
model with $SU(2)$ global symmetry on each site, in which the gauge symmetry 
is $SU(2)$ in the middle site and only $U(1)$ on the boundary 
sites.\footnote{The same arguments can be employed in 5D models that include 
the full electroweak symmetry, for example  in the model of 
Ref.~\cite{Contino:2003ve} with the gauge group $SU(3)\times U(1)$ broken to 
$SU(2)\times U(1)$.} At the very low energy only the 
diagonal $U(1)$ gauge group and the diagonal $SU(2)$ global symmetry are unbroken. The gauge 
sector has in the $U(1)$ sector one massless and two massive modes, as 
discussed in Section 3, as well as massive $W^{\pm}$ gauge bosons 
corresponding to the broken $SU(2)$ generators gauged in the middle site. The 
"Higgs" is taken as the $A_5$ component of the bulk $U(1)$ gauge field.

 In this 
three-site model, one massless and one massive $U(1)$ fields, as well as the 
massive $W^{\pm}$,  are even under KK parity, which is the reflection of the 
two boundary sites. There is also one massive $U(1)$ field that is KK odd. It 
is worth observing that the massive even mode is always heavier than the 
massive odd mode and has a wave function localized near the middle site when 
$g_b>g_a$, as can be seen in Eq.~(\ref{threesiteeig}). In other words, if one 
takes the limit that $g_b\gg g_a$ and $g_b \gg 1$, the massive even mode 
becomes very heavy and should be integrated out of the effective theory, which 
is equivalent to integrating out the middle site. Therefore, one is left with 
a two-site model and only one massive mode that is odd under KK-parity. 
However, the mass of the Higgs, a scalar that corresponds to the $A_5$ 
component of the $U(1)$ field, should still be protected by the 
pseudo-Goldstone (or non-local) nature of the scalar itself. As such, its mass 
must be cut off by the only massive vector boson in the spectrum, i.e. the odd 
KK gauge boson. Alternatively, one could start with the two-site model and 
gradually ``integrate-in'' the other massive modes and the other sites. 
Obviously such UV operations  cannot change the infrared nature of the 
question.

\subsection*{B.2  Full 5D Calculation}

In order to prove our assertion that the Higgs mass  can be cut off by the 
lightest gauge KK mode, we investigate below a simple toy 
gauge-Higgs model and compute the Higgs mass in a fully 5D calculation.  
We first consider a gauge-Higgs model in the usual RS framework with two branes, 
and later we extend the analysis to the KK parity symmetric IR-UV-IR setup. We 
will provide technical arguments showing that the contributions to the Higgs 
mass are indeed cut off by the mass of the {\it lightest} gauge KK mode, which 
is odd under KK parity in our setup. 

We start with the usual 2-brane RS1 setup where the 
 boundary conditions for the three $SU(2)$ generators of  are given by: 
\beq
\label{a3++}
\bvec A_\mu^{1,2} [--] 
\\ A_\mu^3[++]  
\evec 
\eeq  
where $[+]$ stands for the Neumann (or mixed, in the presence of brane kinetic terms),
while $[-]$ stands for the Dirichlet boundary conditions on the UV and IR brane, respectively.
At energies below the compactification scale, we recover a $U(1)$ gauge theory with a charged scalar originating from the $A_5$ component. 
The latter plays the role of the Higgs - its vev  breaks the remaining $U(1)$ symmetry. 

To investigate contributions to the Higgs potential from the gauge fields it 
is convenient to employ the formalism  of Ref.~\cite{aa} which we briefly 
review below.  The Higgs potential is derived from the so-called spectral 
function $\rho(s) \equiv \det (- s  + m_n^2)$, whose zeros on the positive 
real axis encode the whole KK spectrum in the presence of the gauge-Higgs vev 
$\la A_5\ra $. The spectral function can be computed by solving the equations 
of motion  and imposing the boundary conditions. This procedure yields a 
quantization condition for the KK masses, which can be used as the spectral 
function.  With  the spectral function at hand, we can compute the Higgs 
potential from the Coleman-Weinberg formula,   \beq
\label{e.cws} 
V = {N \over 16 \pi^{2}}  \int_0^\infty dp  p^3 \log \left ( \rho(-p^2) \right )  
\eeq 
where $N = +3$ for the gauge fields.

We move to solving the equations of motion. The $SU(2)$ gauge field is 
expanded into KK modes as $A_\mu = A_{\mu,n}(x) f_n(y)$. The profile $f(y)$ is 
an adjoint matrix, $f = f^a \sigma^a$, and it  satisfies the equation of 
motion: \beq D_y (e^{-2 k y} D_y f) + p^2 f = 0  \eeq with $D_y f = \pa_y f - 
i g_5 [\la A_5 \ra, f]$, so that various components $f^a$ are mixed in the 
presence of  the gauge-Higgs vev. We can rotate away the vev from the 
equations of motion by rewriting  \beq \label{e.hr} f(y) = \Omega(y) \hat f 
(y) \Omega^\dagger(y) \qquad \Omega(y) = e^{i g_5 \int_0^y \la A_5 \ra} \eeq  
and we obtain the simple equation:   \beq \label{e.eomgp} \pa_y (e^{-2 k y} 
\pa_y \hat f) + p^2 \hat f = 0  \eeq in which  $\hat f^a$ do not mix with each 
other.  The gauge-Higgs vev is now shifted to  the IR boundary conditions for 
$\hat f$ (the UV boundary conditions are unchanged because we fixed $\Omega(0) 
= 1$). Eq.~(\ref{e.eomgp}) is solved in terms of the Bessel and Neumann 
functions, $a^{-1}(y) Z_1(p/k a(y))$, $Z = J,Y$.  We define two combinations 
of these solutions, $C(y)$ and $S(y)$,  that satisfy $C(0) = 1$,  $C'(0) = 0$, 
$S(0) = 0$, $S'(0) = p$. Using this notation, we can write down the solutions 
to Eq.~(\ref{e.eomgp})  in such a way that the profiles $f(y)$ satisfy the UV 
boundary  conditions: \beq
\label{e.lgp} 
\hat f^{1,2}(y) = \alpha_{1,2} S(y) \qquad \hat f^{3}(y) = \alpha_{3} C(y) 
\qquad \eeq The profiles $f(y)$ are found by rotating $\hat  f$ with  
$\Omega(y)$, as in Eq.~(\ref{e.hr}).  To compute $\Omega$, we choose the Higgs 
vev to reside along $\sigma^1$, \beq \la A_5 \ra _=   {a^{-2}(y) \over \left 
[\int_0^L a^{-2} \right ]^{1/2} } {\sigma^1 \over 2} \ti v  \eeq which leads 
to \beq \Omega(L) = \left ( \ba{cc} \cos(\ti v/2f) & i \sin(\ti v/2f) 
\\
 i \sin(\ti v/2f) & \cos(\ti v/2f)
 \ea \right )
\qquad 
f^2 = {2 k e^{-2 k L} \over g_5^2} 
\eeq 
Hence
\bea
f^1(L) &=& \hat f^1(L) 
\nn
f^2(L) &=&  \cos(\ti v/f) \hat f^2(L)  + \sin (\ti v/f) \hat f^3(L)
\nn
f^3(L) &=&  - \sin(\ti v/f) \hat f^2(L)  + \cos (\ti v/f) \hat f^3(L) 
\eea 
We impose IR boundary conditions (with the IR brane kinetic term for $A_\mu^3$) and we solve the resulting set of equations.  
One solution is $\alpha_{2,3}=0$ and $S(L) = 0$, but here the quantization condition does not depend on the Higgs vev, therefore these eigenstates do not contribute to the Higgs potential. 
The other solution is $\alpha_1 = 0$ and  
\bea
0 &=&  \alpha_2 \cos(\ti v/f) S(L)  + \alpha_3 \sin (\ti v/f) C(L) 
\nn
0
&=&  - \alpha_2  \sin(\ti v/f) \left  ( S'(L) - p^2 r_{IR} a^{-2}(L)  S(L) \right )   
+ \alpha_3 \cos (\ti v/f) \left (  C'(L)  -  p^2 r_{IR} a^{-2}(L)  C(L) \right ) 
\nn
\eea 
The determinant of this set yields the quantization condition, ergo the spectral function
\beq
\label{e.lms} 
\rho_{+}(p^2) = F(p^2) + \sin^2(\ti v/f) \eeq where the form factor is $F(p^2) 
=  p^{-1} a^2(L) C'(L) S(L)  -  p r_{IR}  C(L) S(L)$. For the large BKT, the 
form factor below the KK scale $\mkk = k e^{- k L}$  can be well approximated 
by a simple polynomial in $p^2$, 
\beq 
F(p^2) \approx - {p^2 \over g_0^2 f^2} \left (1 - {p^2 \over m_{1_+}^2} 
\right ) 
\eeq 
where $m_{1_+} \approx (r_{IR}/L)^{1/2} g_0 f$ is the mass of the lightest KK 
mode. This mode will correspond to the lightest {\it even} KK mode, given the $(++)$
boundary condition in the $A_3$ component in Eq.~(\ref{a3++}), when we move to the 
model with KK parity. For $p \gg \mkk$ the form factor grows exponentially,   
$F(p^2) \sim e^{2 p/\mkk}$ for $p \to \infty$, which ensures that the 
gauge-Higgs potential is UV finite. The Higgs mass parameter is given by the 
integral of the inverse form factor,   
\beq 
V_+''(\ti v = 0)  = {N \over 8 
\pi^{2} f^2}  \int_0^\infty dp p^{3} {1  \over F(p^2)}   
\eeq 
The integral is 
dominated by the low energy contribution, and we can estimate 
\beq V_+''(0) 
\sim N/(4\pi^2) m_{1_+}^2 \log(\mkk/m_{1_+}) 
\eeq 
We can see that the Higgs mass 
is cut off by the lightest KK mode, which in the usual RS1 model is a few TeV.


Now, we move to the KK parity symmetric setup.   
In the presence of large BKTs, there is an extra mode which can be below the TeV scale. 
We anticipate that naturalness will be improved and we check it explicitly below.
As discussed in Section \ref{IRUVIR}, 
introducing a mirror AdS slice is equivalent to introducing another gauge field whose UV boundary conditions are flipped with respect to the original one, 
\beq
\bvec 
\ti A_\mu^{1,2} [+-] 
\\ 
\ti A_\mu^3[-+]  
\evec 
\eeq      
The contribution from this extra gauge field needs to be included in the 
Higgs potential.  Repeating the same steps we find the spectral function 
\beq 
\rho_{-}(p^2) = F(p^2) + \cos^2(\ti v/f) 
\eeq  
where the form factor $F(p^2)$ 
is the same as in Eq.~(\ref{e.lms}). Note that, for $\ti v = 0$, $\rho_{-}$ 
has a zero for $p \approx g_0 f$, which implies that the mass of the lightest 
{\em odd} mode is $m_{1_-} \approx g_0 f$.    The KK parity partner of the 
gauge multiplet contributes to the Higgs mass as  
\beq 
V_-''(\ti v = 0)  = -{N 
\over 8 \pi^{2} f^2}  \int_0^\infty dp p^{3} {1  \over (F(p^2) + 1)}   
\eeq   
Summing up the two contributions we have 
\beq 
V''(\ti v = 0) =   {N \over 8 
\pi^{2} f^2}   \int_0^\infty dp p^{3} { 1 \over F(p^2) (F(p^2) + 1)}  
\eeq 
The presence of the KK partner greatly reduces the sensitivity of the Higgs mass 
parameter to the high scales. For $p < m_{1_-}$, the integrand is of order 
$\sim p m_{1_-}^2$, but for  $p >  m_{1_-}$ we have  $F(p^2) > 1$ and the 
integrand switches to softer UV behavior, $\sim  m_{1_-}^4/p$, so that the mass 
is only logarithmically sensitive to momenta above $m_{1_-}$. From that, we can 
estimate 
\beq 
V''(0) \sim N/ (4 \pi^2) m_{1_-}^2 \log(m_{1_+}/m_{1_-}) 
\eeq 
Similar to the RS1 setup without KK parity, the Higgs mass parameter generated 
by loops of the gauge KK modes 
ends up being of the order of the mass of the {\it lightest} KK mode, which in this
case is odd under KK parity.   
Since the odd mode can be lighter than 1 TeV, without conflicting with the  
electroweak precision tests, we can address the little hierarchy problem and 
improve on the naturalness in the KK 
parity symmetric setup. 




\begin{thebibliography}{99}


\bibitem{Barbieri:1999tm}
  R.~Barbieri and A.~Strumia,
  Phys.\ Lett.\  B {\bf 462}, 144 (1999)
  [arXiv:hep-ph/9905281];
%
  R.~Barbieri and A.~Strumia,
  arXiv:hep-ph/0007265;
%
  R.~Barbieri, A.~Pomarol, R.~Rattazzi and A.~Strumia,
  Nucl.\ Phys.\  B {\bf 703}, 127 (2004)
  [arXiv:hep-ph/0405040];
%
  Z.~Han and W.~Skiba,
  Phys.\ Rev.\  D {\bf 71}, 075009 (2005)
  [arXiv:hep-ph/0412166];
  G.~Cacciapaglia, C.~Csaki, G.~Marandella and A.~Strumia,
  Phys.\ Rev.\  D {\bf 74}, 033011 (2006)
  [arXiv:hep-ph/0604111].




\bibitem{Peskin:1991sw}
  M.~E.~Peskin and T.~Takeuchi,
  Phys.\ Rev.\  D {\bf 46}, 381 (1992);
  M.~Golden and L.~Randall,
  Nucl.\ Phys.\  B {\bf 361}, 3 (1991).



\bibitem{Wudka:2003se}
  J.~Wudka,
  arXiv:hep-ph/0307339.

\bibitem{Cheng:2003ju}
  H.~C.~Cheng and I.~Low,
  JHEP {\bf 0309}, 051 (2003)
  [arXiv:hep-ph/0308199].


%
\bibitem{Cheng:2004yc}
  H.~C.~Cheng and I.~Low,
  JHEP {\bf 0408}, 061 (2004)
  [arXiv:hep-ph/0405243];
%
  I.~Low,
  JHEP {\bf 0410}, 067 (2004)
  [arXiv:hep-ph/0409025].


\bibitem{Cheng:2002ab}
 H.~C.~Cheng, K.~T.~Matchev and M.~Schmaltz,
Phys.\ Rev.\  D {\bf 66}, 036005 (2002)
[arXiv:hep-ph/0204342].


\bibitem{Appelquist:2000nn}
  T.~Appelquist, H.~C.~Cheng and B.~A.~Dobrescu,
  Phys.\ Rev.\  D {\bf 64}, 035002 (2001)
  [arXiv:hep-ph/0012100].



\bibitem{Randall:1999ee}
L.~Randall and R.~Sundrum,
Phys.\ Rev.\ Lett.\  {\bf 83}, 3370 (1999)
[arXiv:hep-ph/9905221].



\bibitem{Maldacena:1997re}
  J.~M.~Maldacena,
  Adv.\ Theor.\ Math.\ Phys.\  {\bf 2}, 231 (1998)
  [Int.\ J.\ Theor.\ Phys.\  {\bf 38}, 1113 (1999)]
  [arXiv:hep-th/9711200];
%
  S.~S.~Gubser, I.~R.~Klebanov and A.~M.~Polyakov,
  Phys.\ Lett.\ B {\bf 428}, 105 (1998)
  [arXiv:hep-th/9802109].
%

\bibitem{Witten:1998qj}
  E.~Witten,
  Adv.\ Theor.\ Math.\ Phys.\  {\bf 2}, 253 (1998)
  [arXiv:hep-th/9802150].


\bibitem{Arkani-Hamed:2000ds}
  N.~Arkani-Hamed, M.~Porrati and L.~Randall,
  JHEP {\bf 0108}, 017 (2001)
  [arXiv:hep-th/0012148].

\bibitem{Rattazzi:2000hs}
  R.~Rattazzi and A.~Zaffaroni,
  JHEP {\bf 0104}, 021 (2001)
  [arXiv:hep-th/0012248].

\bibitem{Contino:2003ve}
  R.~Contino, Y.~Nomura and A.~Pomarol,
  Nucl.\ Phys.\ B {\bf 671}, 148 (2003)
  [arXiv:hep-ph/0306259].



\bibitem{gn}Y.~Grossman and M.~Neubert,
Phys.\ Lett.\ B {\bf 474}, 361 (2000)
[arXiv:hep-ph/9912408].

\bibitem{gp}
T.~Gherghetta and A.~Pomarol,
Nucl.\ Phys.\ B {\bf 586}, 141 (2000)
[arXiv:hep-ph/0003129].


\bibitem{bulkgauge}
  H.~Davoudiasl, J.~L.~Hewett and T.~G.~Rizzo,
  Phys.\ Lett.\ B {\bf 473}, 43 (2000)
  [arXiv:hep-ph/9911262];
%
  A.~Pomarol,
  Phys.\ Lett.\ B {\bf 486}, 153 (2000)
  [arXiv:hep-ph/9911294].
%
  S.~Chang, J.~Hisano, H.~Nakano, N.~Okada and M.~Yamaguchi,
  Phys.\ Rev.\  D {\bf 62}, 084025 (2000)
  [arXiv:hep-ph/9912498].

\bibitem{Huber:2000ie}
S.~J.~Huber and Q.~Shafi,
Phys.\ Lett.\ B {\bf 498}, 256 (2001)
[arXiv:hep-ph/0010195].

\bibitem{Agashe:2004cp}
K.~Agashe, G.~Perez and A.~Soni,
Phys.\ Rev.\ Lett.\  {\bf 93}, 201804 (2004)
[arXiv:hep-ph/0406101];
Phys.\ Rev.\ D {\bf 71}, 016002 (2005)
[arXiv:hep-ph/0408134].

\bibitem{NMFV}
K.~Agashe, M.~Papucci, G.~Perez and D.~Pirjol,
  arXiv:hep-ph/0509117;
  Z.~Ligeti, M.~Papucci and G.~Perez,
  Phys.\ Rev.\ Lett.\  {\bf 97}, 101801 (2006)
  [arXiv:hep-ph/0604112].

\bibitem{Agashe:2003zs}
K.~Agashe, A.~Delgado, M.~J.~May and R.~Sundrum,
JHEP {\bf 0308}, 050 (2003)
[arXiv:hep-ph/0308036];
%


\bibitem{Agashe:2006at}
  K.~Agashe, R.~Contino, L.~Da Rold and A.~Pomarol,
  Phys.\ Lett.\  B {\bf 641} (2006) 62
  [arXiv:hep-ph/0605341];
%
  M.~Carena, E.~Ponton, J.~Santiago and C.~E.~M.~Wagner,
  Nucl.\ Phys.\ B {\bf 759}, 202 (2006)
  [arXiv:hep-ph/0607106] and
%
Phys.\ Rev.\  D {\bf 76}, 035006 (2007)
[arXiv:hep-ph/0701055];
%
  R.~Contino, L.~Da Rold and A.~Pomarol,
  Phys.\ Rev.\  D {\bf 75}, 055014 (2007)
  [arXiv:hep-ph/0612048];
  A.~D.~Medina, N.~R.~Shah and C.~E.~M.~Wagner,
  arXiv:0706.1281 [hep-ph].




\bibitem{others1} For studies with $\sim 10$ TeV KK masses, see 
S.~J.~Huber,
Nucl.\ Phys.\ B {\bf 666}, 269 (2003)
[arXiv:hep-ph/0303183];
%
  S.~Khalil and R.~Mohapatra,
  Nucl.\ Phys.\ B {\bf 695}, 313 (2004)
  [arXiv:hep-ph/0402225].


\bibitem{others2}
  G.~Burdman,
  Phys.\ Lett.\ B {\bf 590}, 86 (2004)
  [arXiv:hep-ph/0310144];
%
  G.~Moreau and J.~I.~Silva-Marcos,
  JHEP {\bf 0603}, 090 (2006)
  [arXiv:hep-ph/0602155];
%
  K.~Agashe, A.~E.~Blechman and F.~Petriello,
  arXiv:hep-ph/0606021.

\bibitem{Cacciapaglia:2007fw}
For a mechanism to suppress FCNC in a model
which does not address flavor hierachy, see
  G.~Cacciapaglia, C.~Csaki, J.~Galloway, G.~Marandella, J.~Terning and A.~Weiler,
  arXiv:0709.1714 [hep-ph].


\bibitem{kkgluon}
For some recent LHC studies of
direct production of KK modes in
this framework, see
  K.~Agashe, A.~Belyaev, T.~Krupovnickas, G.~Perez and J.~Virzi,
  arXiv:hep-ph/0612015;
  A.~L.~Fitzpatrick, J.~Kaplan, L.~Randall and L.~T.~Wang,
  JHEP {\bf 0709}, 013 (2007)
  [arXiv:hep-ph/0701150];
  C.~Dennis, M.~Karagoz Unel, G.~Servant and J.~Tseng,
  arXiv:hep-ph/0701158;
  B.~Lillie, L.~Randall and L.~T.~Wang,
  JHEP {\bf 0709}, 074 (2007)
  [arXiv:hep-ph/0701166];
  K.~Agashe, H.~Davoudiasl, G.~Perez and A.~Soni,
  Phys.\ Rev.\  D {\bf 76}, 036006 (2007)
  [arXiv:hep-ph/0701186];
  F.~Ledroit, G.~Moreau and J.~Morel,
  JHEP {\bf 0709}, 071 (2007)
  [arXiv:hep-ph/0703262];
%
  B.~Lillie, J.~Shu and T.~M.~P.~Tait,
  arXiv:0706.3960 [hep-ph];
  A.~Djouadi, G.~Moreau and R.~K.~Singh,
  arXiv:0706.4191 [hep-ph];
  K.~Agashe {\it et al.},
  arXiv:0709.0007 [hep-ph];
%
%
%
%
  O.~Antipin, D.~Atwood and A.~Soni,
  arXiv:0711.3175 [hep-ph];
%
%
%
  M.~Carena, A.~D.~Medina, B.~Panes, N.~R.~Shah and C.~E.~M.~Wagner,
  arXiv:0712.0095 [hep-ph].
%
For an overview, see
%
  H.~Davoudiasl, T.~G.~Rizzo and A.~Soni,
  arXiv:0710.2078 [hep-ph].

\bibitem{Servant:2002aq}
  G.~Servant and T.~M.~P.~Tait,
  Nucl.\ Phys.\  B {\bf 650}, 391 (2003)
  [arXiv:hep-ph/0206071].
%

\bibitem{Cheng:2002ej}
  H.~C.~Cheng, J.~L.~Feng and K.~T.~Matchev,
  Phys.\ Rev.\ Lett.\  {\bf 89}, 211301 (2002)
  [arXiv:hep-ph/0207125].


\bibitem{Burdman:2006jj}
  G.~Burdman and A.~G.~Dias,
  JHEP {\bf 0701}, 041 (2007)
  [arXiv:hep-ph/0609181].


\bibitem{Arkani-Hamed:2001ca}
N.~Arkani-Hamed, A.~G.~Cohen and H.~Georgi,
Phys.\ Rev.\ Lett.\  {\bf 86}, 4757 (2001)
[arXiv:hep-th/0104005].

\bibitem{Hill:2000mu}
C.~T.~Hill, S.~Pokorski and J.~Wang,
Phys.\ Rev.\  D {\bf 64}, 105005 (2001)
[arXiv:hep-th/0104035].


\bibitem{Cacciapaglia:2005pa}
  G.~Cacciapaglia, C.~Csaki, C.~Grojean, M.~Reece and J.~Terning,
  Phys.\ Rev.\  D {\bf 72}, 095018 (2005)
  [arXiv:hep-ph/0505001].

\bibitem{Cacciapaglia:2006tg}
  G.~Cacciapaglia, C.~Csaki, C.~Grojean and J.~Terning,
  Phys.\ Rev.\  D {\bf 74}, 045019 (2006)
  [arXiv:hep-ph/0604218].


\bibitem{Goldberger:1999uk}
  W.~D.~Goldberger and M.~B.~Wise,
  Phys.\ Rev.\ Lett.\  {\bf 83}, 4922 (1999)
  [arXiv:hep-ph/9907447];
%
  J.~Garriga and A.~Pomarol,
  Phys.\ Lett.\  B {\bf 560}, 91 (2003)
  [arXiv:hep-th/0212227].


\bibitem{Hill:2007zv}
  C.~T.~Hill and R.~J.~Hill,
  arXiv:0705.0697 [hep-ph].

\bibitem{ArkaniHamed:2001is}
  N.~Arkani-Hamed, A.~G.~Cohen and H.~Georgi,
  Phys.\ Lett.\  B {\bf 516}, 395 (2001)
  [arXiv:hep-th/0103135].


\bibitem{Carena:2002dz}
  M.~Carena, E.~Ponton, T.~M.~P.~Tait and C.~E.~M.~Wagner,
  Phys.\ Rev.\  D {\bf 67}, 096006 (2003)
  [arXiv:hep-ph/0212307].

\bibitem{Davoudiasl:2002ua}
  H.~Davoudiasl, J.~L.~Hewett and T.~G.~Rizzo,
  Phys.\ Rev.\  D {\bf 68}, 045002 (2003)
  [arXiv:hep-ph/0212279].



\bibitem{holoRG} For applications to compact slice
of AdS, see, for example, 
  A.~Lewandowski, M.~J.~May and R.~Sundrum,
  Phys.\ Rev.\  D {\bf 67}, 024036 (2003)
  [arXiv:hep-th/0209050];
  A.~Lewandowski and M.~Redi,
  Phys.\ Rev.\  D {\bf 68}, 044012 (2003)
  [arXiv:hep-th/0305013];
  A.~Lewandowski,
  Phys.\ Rev.\  D {\bf 71}, 024006 (2005)
  [arXiv:hep-th/0409192].


\bibitem{higgsless}
C.~Csaki, C.~Grojean, L.~Pilo and J.~Terning,
Phys.\ Rev.\ Lett.\  {\bf 92}, 101802 (2004)
[arXiv:hep-ph/0308038].

\bibitem{Kaplan:2001ga}
D.~E.~Kaplan and T.~M.~P.~Tait,
JHEP {\bf 0111}, 051 (2001)
[arXiv:hep-ph/0110126];
H.~Georgi, A.~K.~Grant and G.~Hailu,
Phys.\ Rev.\  D {\bf 63}, 064027 (2001)  [arXiv:hep-ph/0007350].


\bibitem{Agashe:2004ci}
  K.~Agashe and G.~Servant,
  Phys.\ Rev.\ Lett.\  {\bf 93}, 231805 (2004)
  [arXiv:hep-ph/0403143] and
  JCAP {\bf 0502}, 002 (2005)
  [arXiv:hep-ph/0411254].

\bibitem{Belanger:2007dx}
  G.~Belanger, A.~Pukhov and G.~Servant,
  arXiv:0706.0526 [hep-ph].

\bibitem{Cheng:2002iz}
  H.~C.~Cheng, K.~T.~Matchev and M.~Schmaltz,
  Phys.\ Rev.\  D {\bf 66}, 036005 (2002)
  [arXiv:hep-ph/0204342].

\bibitem{Burnell:2005hm}
  F.~Burnell and G.~D.~Kribs,
  Phys.\ Rev.\  D {\bf 73}, 015001 (2006)
  [arXiv:hep-ph/0509118].
  K.~Kong and K.~T.~Matchev,
  JHEP {\bf 0601}, 038 (2006)
  [arXiv:hep-ph/0509119].


\bibitem{Agashe:2004rs}
  K.~Agashe, R.~Contino and A.~Pomarol,
  Nucl.\ Phys.\  B {\bf 719}, 165 (2005)
  [arXiv:hep-ph/0412089];
%
  K.~Agashe and R.~Contino,
  Nucl.\ Phys.\  B {\bf 742}, 59 (2006)
  [arXiv:hep-ph/0510164].


\bibitem{Servant:2002hb}
  G.~Servant and T.~M.~P.~Tait,
  New J.\ Phys.\  {\bf 4}, 99 (2002)
  [arXiv:hep-ph/0209262].


%
\bibitem{Dvali:2000hr}
  G.~R.~Dvali, G.~Gabadadze and M.~Porrati,
  Phys.\ Lett.\  B {\bf 485}, 208 (2000)
  [arXiv:hep-th/0005016].


\bibitem{Luty:2003vm}
M.~A.~Luty, M.~Porrati and R.~Rattazzi,
JHEP {\bf 0309}, 029 (2003)
[arXiv:hep-th/0303116].


\bibitem{Low:2000pq}
I.~Low and A.~Zee,
Nucl.\ Phys.\  B {\bf 585}, 395 (2000)
[arXiv:hep-th/0004124].

\bibitem{Kanti:2000rd}
  P.~Kanti, K.~A.~Olive and M.~Pospelov,
  Phys.\ Lett.\  B {\bf 481}, 386 (2000)
  [arXiv:hep-ph/0002229].

\bibitem{Mukohyama:2000wq}
  S.~Mukohyama,
  Phys.\ Rev.\  D {\bf 63}, 044008 (2001)
  [arXiv:hep-th/0007239].

\bibitem{Hofmann:2000cj}
  R.~Hofmann, P.~Kanti and M.~Pospelov,
  Phys.\ Rev.\  D {\bf 63}, 124020 (2001)
  [arXiv:hep-ph/0012213].

\bibitem{Freedman:1999gp}
D.~Z.~Freedman, S.~S.~Gubser, K.~Pilch and N.~P.~Warner,
Adv.\ Theor.\ Math.\ Phys.\  {\bf 3}, 363 (1999)
[arXiv:hep-th/9904017].

\bibitem{van Dam:1970vg}
H.~van Dam and M.~J.~G.~Veltman,
Nucl.\ Phys.\  B {\bf 22}, 397 (1970);
V.~I.~Zakharov,
JETP Lett.\  {\bf 12} (1970) 312

\bibitem{ArkaniHamed:2002sp}
N.~Arkani-Hamed, H.~Georgi and M.~D.~Schwartz,
Annals Phys.\  {\bf 305}, 96 (2003)
[arXiv:hep-th/0210184].

\bibitem{Hosotani:1983xw}
Y.~Hosotani,
Phys.\ Lett.\  B {\bf 126}, 309 (1983).


\bibitem{ArkaniHamed:2001nc}
N.~Arkani-Hamed, A.~G.~Cohen and H.~Georgi,
Phys.\ Lett.\  B {\bf 513}, 232 (2001)
[arXiv:hep-ph/0105239].

\bibitem{Contino:2004vy}
  R.~Contino and A.~Pomarol,
  JHEP {\bf 0411}, 058 (2004)
  [arXiv:hep-th/0406257].

\bibitem{aa}
  A.~Falkowski,
  Phys.\ Rev.\ D {\bf 75}, 025017 (2007)
  [arXiv:hep-ph/0610336]. 
  A.~Falkowski,
  arXiv:0710.4050 [hep-ph].



%







\end{thebibliography}
\end{document}